\documentclass[3p,review,sort&compress]{elsarticle}

\usepackage[svgnames]{xcolor}
\usepackage{graphicx}      
\usepackage{natbib}
\usepackage{amsmath}
\usepackage{bm}
\usepackage{amssymb}
\usepackage{threeparttable}
\usepackage{multirow}
\usepackage{ucs}
\usepackage[utf8x]{inputenc}
\usepackage{mathtools}
\usepackage[T1]{fontenc}
\usepackage{engord}
\usepackage[subrefformat=parens,labelformat=parens]{subfig}
\usepackage[inline]{enumitem}
\usepackage{booktabs}
\usepackage{mathrsfs}
\usepackage[colorinlistoftodos]{todonotes}
\usepackage[acronym]{glossaries}
\usepackage{subfig}
\usepackage{xfrac}
\usepackage{engord}
\usepackage{enumitem}
\usepackage[version=4]{mhchem}
\usepackage[section]{placeins}
\usepackage{siunitx}
\usepackage{arydshln}
\usepackage{url}
\usepackage{circuitikz}

\usepackage{tabularx} 
\usetikzlibrary{positioning,shapes,shadows,arrows,decorations.pathreplacing,er,shapes.multipart}
\usetikzlibrary{arrows, decorations.markings}
\usetikzlibrary{matrix,decorations.pathmorphing,positioning}
\usetikzlibrary{fit}
\usetikzlibrary{calc}
\usetikzlibrary{arrows}

\usepackage{tikz}
\usepackage{circuitikz}
\usepackage{url}
\usepackage{siunitx}

\usepackage{pgfplots}
\pgfplotsset{compat=newest}
\usetikzlibrary{plotmarks}
\usetikzlibrary{arrows.meta}
\usepgfplotslibrary{patchplots}
\usepackage{grffile}
\usepackage{amsmath}

\usepackage{glossaries}
\newacronym{kl}{KL}{Kullback-Leibler}
\newacronym{vb}{VB}{variational Bayes}
\newacronym{fos}{FOS}{fractional order system}
\newacronym{svi}{SVI}{stochastic variational inference}
\newacronym{elbo}{ELBO}{evidence lower bound}
\newacronym{eis}{EIS}{electrochemical impedance spectroscopy}
\newacronym{mcmc}{MCMC}{Markov chain Monte Carlo}
\newacronym{sofc}{SOFC}{solid oxide fuel cell}
\newacronym{ecm}{ECM}{equivalent circuit model}
\newacronym{ecms}{ECMs}{equivalent circuit models}
\newacronym{drt}{DRT}{distribution of relaxation times}
\newacronym{adam}{ADAM}{adaptive moment estimation}
\newacronym{cwt}{ADAM}{continuous wavelet transform}
\newacronym{nuts}{NUTS}{No-U-Turn sampling}
\newacronym{hmc}{HMC}{Hamiltonian Monte Carlo algorithm}
\newacronym{bop}{BoP}{balance of plant}

\usepackage[normalem]{ulem}

\newcommand{\mnewn}[2]{{#2}}
\newcommand{\mnew}[2]{{#2}}
\tikzstyle{startstop} = [circle, rounded corners, minimum width=1cm, minimum height=0.5cm,text centered, draw=black, fill=red!30]
\tikzstyle{empty} = [rectangle, rounded corners, minimum width=0.5cm, minimum height=0.5cm,text centered, draw=black, fill=black!10]
\tikzstyle{process} = [rectangle, minimum width=1cm, minimum height=0.5cm, text centered, draw=black, fill=orange!30]
\tikzstyle{decision} = [rectangle, rounded corners, minimum width=0.5cm, minimum height=0.5cm, text centered, draw=black, fill=green!30]
\tikzstyle{io} = [trapezium, trapezium left angle=70, trapezium right angle=110, text centered, draw=black, fill=blue!30]
\tikzstyle{arrow} = [thick,->,>=stealth]

\makeatletter
\def\TikzBipolePath#1#2{\pgf@circ@bipole@path{#1}{#2}}
\makeatother
\newlength{\ResUp}
\newlength{\ResDown}
\newlength{\ResLeft}
\newlength{\ResRight}

\ctikzset{bipoles/newcomponent/height/.initial=.50}   
\ctikzset{bipoles/newcomponent/width/.initial=.20}    
\pgfcircdeclarebipole{}                               
{\ctikzvalof{bipoles/newcomponent/height}}
{newcomponent}                                        
{\ctikzvalof{bipoles/newcomponent/height}}
{\ctikzvalof{bipoles/newcomponent/width}}
{                                                     
	\pgfsetlinewidth{\pgfkeysvalueof{/tikz/circuitikz/bipoles/thickness}\pgfstartlinewidth}
	\pgfsetlinewidth{\pgfkeysvalueof{/tikz/circuitikz/bipoles/thickness}\pgfstartlinewidth}
	\pgfextractx{\ResRight}{\northeast}
	\pgfextracty{\ResUp}{\northeast}
	\pgfextractx{\ResLeft}{\southwest}
	\pgfextracty{\ResDown}{\southwest}
	\pgfpathmoveto{\pgfpoint{2\ResLeft}{\ResUp}}      
	\pgfpathlineto{\pgfpoint{\ResLeft}{0}}
	\pgfpathlineto{\pgfpoint{2\ResLeft}{\ResDown}}
	\pgfpathmoveto{\pgfpoint{.25\ResLeft}{\ResUp}}      
	\pgfpathlineto{\pgfpoint{\ResRight}{0}}
	\pgfpathlineto{\pgfpoint{0}{\ResDown}}
	\pgfusepath{draw}                                   
}
\def\circlepath#1{\TikzBipolePath{newcomponent}{#1}}
\tikzset{newcomponent/.style = {\circuitikzbasekey, /tikz/to path=\circlepath, l=#1}}

\begin{document}
	\engordraisetrue
	\begin{frontmatter}
		\title{Evaluating uncertainties in electrochemical impedance spectra of solid oxide fuel cells}
		\author[add1,mps]{Luka Žnidarič\corref{cor1}}
		\ead{luka.znidaric@ijs.si}
		\author[add1,mps]{Gjorgji Nusev}
		\author[cea]{Bertrand Morel}
		\author[cea]{Julie Mougin}
		
		\author[add1]{Đani Juričić}
		\author[add1]{Pavle Boškoski}
		\address[add1]{Jožef Stefan Institute, Jamova cesta 39, SI-1000 Ljubljana, Slovenia}
		\address[mps]{Jožef Stefan International Postgraduate School, Jamova cesta 39, SI-1000 Ljubljana, Slovenia}
		\address[cea]{Laboratoire des Technologies Hydrogèn
			Commissariat à l’\'Energie Atomique et aux \'Energies Alternatives - CEA,
			CEA/LITEN/DTBH/STHB/LTH
			17 rue des Martyrs – 38054 GRENOBLE CEDEX 9, France
		}
		\cortext[cor1]{Corresponding author.}

		\begin{abstract}   
			\Gls{eis} is a widely used tool for characterization of fuel cells and other electrochemical conversion systems. 
			When applied to the on-line monitoring in the context of in-field applications, the  disturbances, drifts and sensor noise may cause severe distortions in the evaluated spectra, especially in the low-frequency part. 
			Failure to ignore the random effects can result in misinterpreted spectra and, consequently, in misleading diagnostic reasoning.
			This fact has not been often addressed  in the research so far. 
			In this paper, we propose an  approach to the quantification of the spectral uncertainty, which relies on evaluating  the uncertainty of the \gls{ecm}. We apply the computationally efficient \gls{vb} method and compare the quality of the results with those obtained with the \gls{mcmc} algorithm. 
			Namely, \gls{mcmc} algorithm returns accurate distributions of the estimated model parameters, while \gls{vb} approach provides the approximate distributions.
			By using simulated and real data we show that
			approximate results provided by \gls{vb} approach,  although slightly over-optimistic, are still  close to the more realistic \gls{mcmc} estimates. 
			A great advantage of the \gls{vb} method for online monitoring is low computational load, which is several orders of magnitude lower compared to \gls{mcmc}.
			The performance of \gls{vb} algorithm is demonstrated on a case of \gls{ecm} parameters estimation in
			a 6 cell \gls{sofc} stack.
			The complete numerical implementation for recreating the results can be found at \url{https://repo.ijs.si/lznidaric/variational-bayes-supplementary-material}.
		\end{abstract}
		
		
	\end{frontmatter}
	\glsresetall



\section{Introduction}

Currently available tools for characterising electrochemical energy systems predominantly rely on Nyquist curves obtained through \gls{eis}~\cite{EIS2014}.
Conceptually simple and well understood, \gls{eis} analysis has become a standard tool for characterising the health condition of cells and stacks. 
Correct evaluation of the \gls{eis} characteristic is subject to several requirements. 

First, the perturbation signal must have low amplitude in order  not to excite the nonlinear modes of the cells dynamics. 
If the amplitudes are too small, there is a risk that the noise-to-signal ratio will decrease, which will reflect in an increased variance of the \gls{eis} estimates. 
Second, the cells should operate in stationary, stable, and repeatable test conditions. This is mandatory for a correct evaluation of the \gls{eis} curves. 
The internal conditions should remain constant during the perturbation session and  no external disturbance should corrupt the measurements. 
Under stable laboratory conditions, the effect of disturbances on stack  are minimised, which usually results in smooth Nyquist curves.
Correctly evaluated \gls{eis} spectra ensures accurate  detection of fault and stack degradation. Detection is done  by checking their similarity  to the reference  \gls{eis} spectra obtained in the healthy state. 
An anomaly in the online spectra might be ascribed to the fault instead to the disturbances, which results in false fault alarm.
Third, to obtain comparable results, measurements on a cell or a stack should be performed at the same operating point. 
The above requirements are not easily met in applications outside laboratories. 

\mnewn{Reviewer 2, Comment 2.6}{In the in-field operation, the presence of disturbances  in the \gls{bop} and system environment is  likely to come into play. In the laboratory conditions there are  opportunities to apply expensive instrumentation (e.g. on-line gas analyser), high-performance actuators, power electronics  and measurement devices. In commercial applications,  a trade-off between performance and cost of implementation should normally be sought. 
	Cost optimization dictates reduction of the number and types of implemented sensors to minimum and use of standard industrial modules for HW realisation. 
	Variations in gas channel,  drift and noise in flows and temperatures can affect the low-frequency part of the Nyquist curve and can lead to  non-smooth results. Increased noise in current and voltage measurements can also contribute to the non-smooth \gls{eis} evaluation. 
	In addition, improper operating conditions, such as increased fuel utilization, can cause dispersed values of impedance in the low-frequency part of the spectra. The reason is the local fuel starvation, which then affects the voltage variance.}

\begin{figure}
	\centering
	\includegraphics[width=\linewidth]{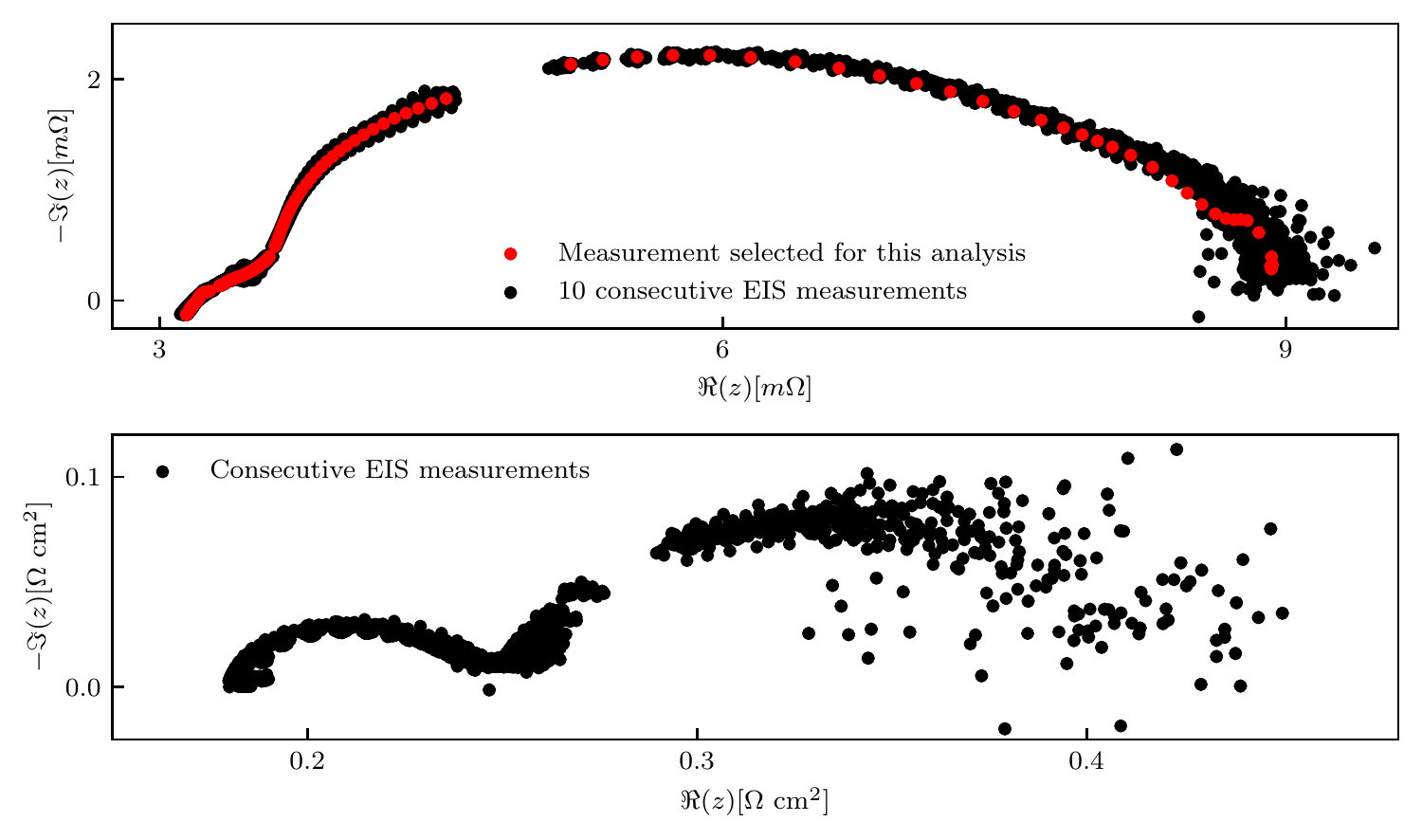}
	\caption{Evaluated \gls{eis} characteristics in laboratory conditions (above) and during an in-field application (below).}
	\label{fig:consecutiveeis}
\end{figure}

\mnewn{Reviewer 1, Comment 1.1, Reviewer 2, Comment 2.3}{
	A strong motivation for the work stems on one hand from experience with some cases of practical implementations and, on the other hand, by the fact that it has not been explicitly  addressed in the literature.
	For better illustration, \figurename~\ref{fig:consecutiveeis} shows two sets of evaluated \gls{eis} curves obtained from successive measurements.
	The first set was obtained on a short \gls{sofc} stack consisting of six planar anode-supported cells installed in an electrical furnace. 
	The stack was operated in the laboratory conditions under nominal current of 32~\si{\ampere} and FU=77.5~\%. More details can be found in \cite{NUSEV2021229491}.
	The second case is an example of a commercial stack with anode-supported cells  operated at around $750^o$C and at the power  $1$kW.
	As fuel,  natural gas is used in a way that it is first partly transformed to syngas by steam reforming.
	During the long-term run there were occasional issues related to the  periodic component in the fuel and steam flow, which was attributed to the control system issues.
	Note that  perturbations do not significantly affect the high frequency range arcs in the \gls{eis} curve.}

It seems the problem is relevant for the \gls{sofc} systems domain. 
With the increasing need for automated condition monitoring, robust solutions are mandatory to ensure reliable  diagnosis. 
This is important since the mitigation actions are taken based on the diagnostic results. 
Unreliable \gls{eis} evaluations can result in misleading diagnosis and countermeasures that can even worsen the condition of a cell or a stack.  

An idea of how to deal with the problem in \glspl{sofc} was recently proposed in articles~\cite{Mougin2019,GALLO2020115718}. 
The authors perform inference on the \gls{ecm} based on a combination of \gls{eis} data smoothing and \gls{eis} averaging. 
Instead of using a single \gls{eis} measurement, the data from several consecutive \gls{eis} measurements are averaged to reduce the dispersion of the \gls{eis} estimates in the low-frequency region. 
This solution, while simple, requires a certain number of measurements spanned on a time window to reliably assess the change in  \gls{eis} characteristic.

In this paper, we propose a novel approach, not pursued before, which is able to account for the influence of perturbations on the evaluated \gls{eis} characteristic, and consequently on the equivalent circuit model, by a statistical modelling approach. 
Properly quantified uncertainties lead to diagnostic solutions that, rather than making a fault statement in clear yes/no categories, suggest the probability that a particular fault is present~\cite{RAKAR1999555}. 
This is essential for cautious and more reliable diagnosis of \glspl{sofc}, which can be additionally enhanced with the operator's prior knowledge.

In the area of electrochemical energy conversion systems, there have been only limited attempts to analyze the stochastic nature of the parameters. 
\mnewn{Reviewer 2, Comment 2.8}{Since parameters of the ECM model are usually obtained by constrained nonlinear optimization, the uncertainty of the estimated parameters is obtained from the local properties of the objective function around the optimal parameters, c.f.~\cite{aa2621ddfa01435aaab80bdd75ff07ad, leonide2010sofc}.
	The problem with the approach is that the approximation is rather rough and might suggest a misleading uncertainty region~\cite{wang2019variational}.} 
To obtain a realistic estimate of the parameters uncertainty, more elaborated approaches should be applied.
Of the available methods, the \gls{mcmc} ~\cite{7873246} provides the most accurate probability distributions for estimated model parameters.
The key problem with the approach is the overwhelming  computational time, which increases rapidly with the number of unknown parameters. 
That renders the approach inappropriate for in-field on-line condition monitoring. 
Instead, a computationally less excessive but still sufficiently accurate method is desired. 

A remedy for the issues imposed by \gls{mcmc} is the  \gls{vb}~\cite{Smidl2006a}. 
Unlike the \gls{mcmc} algorithm that results in the ``true'' posterior distribution, the \gls{vb} approach provides the closest approximation of the posterior distribution using the probability distribution functions from the family of exponential functions.
Consequently, instead of solving the typically intractable evidence integral in pure Bayesian approach, the posterior distribution in \gls{vb} algorithm is found by means of optimization.
The result contains inherent bias, which is  a low price to pay considering the impressive computational efficiency of the approach even for multidimensional cases.
\Gls{vb} approach has been extensively applied in various areas, such as Gaussian process modeling~\cite{Bui2016, Hensman2014}, deep generative models~\cite{Rezende2014}, compressed sensing~\cite{Yang2012,Oikonomou2019}, Hidden Markov models~\cite{Gruhl2016, Panousis2020}, reinforcement learning and control~\cite{Levine2018}. 
\Gls{vb} approach is also applicable to energy management problems~\cite{Liu2020,Liu2019,Choi2018,Choi2018a,Pasquier2020}.

In this paper we  analyse the nature  of the uncertainty regions for the ECM parameters  and compare  the results of the computationally feasible VB approach with the results of \gls{mcmc} approach.   
To the best of the authors' knowledge, the first and only attempt to study the  uncertainty of \gls{ecm} estimates has been done in \cite{electrimacs}. 
That work is the first report on the probability distribution functions of the ECM parameters under nominal operating conditions. 
Unfortunately, the method is  infeasible for on-line monitoring since it takes several hours even with a strong HPC infrastructure.\footnote{HPC Meister, capability: $244$ TFLOPs} 
For example, 25 hours are needed to obtain the correct distributions of parameters for the experimental measurement presented later on. 
What we suggest below is to approximate the true posterior probability density function with a closed form distribution that best captures the nature of the true posterior.
The benefits of having  the uncertainties of the model parameters explained in the closed form are twofold.
First, we get a useful indicator of the quality of the selected ECM model structure and the quality of the experimental data leading to the model. 
Second, it becomes possible to rather easily employ statistical reasoning tools for detecting changes in the  ECM parameters and thus the  deterioration of the system behaviour.
That is the first such approach in the domain of \gls{sofc}.

\mnewn{}{The organisation of the paper is as follows.  
	A brief introduction to the \gls{mcmc} and \gls{vb} approach is given in section~\ref{sec:method}.
	The performance of the method in terms of computational efficiency and accuracy is first demonstrated on a simulated \gls{ecm} in section~\ref{sec:numeric}.
	Finally, the \gls{vb} algorithm is applied to the identification of \gls{ecm} parameters based on data obtained on a \gls{sofc} in section~\ref{sec:results}.
	The main body of the paper is followed by 3 appendices. 
	In~\ref{supp_mat} the complete numerical implementation for recreating the results is described.
	Additional results on \gls{vb} on  simulated measurements with varying degrees of noise can be found in~\ref{Additional_numerical}.
	Finally, some results of the \gls{vb} approach on experimental measurements under different conditions are delivered in~\ref{Additional_experimental}.}

\section{Methodology}
\label{sec:method}
Assume we want to describe a process with a model $\mathcal{M}_{\bm{\theta}}$ parameterized by a vector $\bm{\theta}$. Based on measurements $\bm{\mathbf{x}}$ obtained from the  process, one would like to get  the unknown $\bm{\theta}$ from $\bm{\mathbf{x}}$. Assume also there is some prior knowledge (or guess) about $\bm{\theta}$ expressed in terms of a probability distribution (also called prior distribution\footnote{A prior probability distribution of an uncertain parameter $\bm{\theta}$ is the probability distribution that reflect  one's belief about the parameter before  evidence in terms of data is taken into account.} ) $p(\bm{\theta})$. Prior knowledge can be updated (improved) with information contained in the data by means of Bayes rule. The result is the posterior distribution\footnote{Posterior probability is the updated probability for the event, after taking into account the prior knowledge and information contained in the data.} $p(\bm{\theta}|\bm{\mathbf{x}})$ as follows
\begin{equation}\label{eq:BT}
	\underbrace{p(\bm{\theta}|\mathbf{x})}_{\text{Posterior}} = \frac{\overbrace{p(\mathbf{x}|\bm{\theta})}^{\text{Likelihood}}\overbrace{p(\bm{\theta})}^{\text{Prior}}}{\underbrace{p(\mathbf{x})}_{\text{Evidence}}} 
\end{equation}
The likelihood can be calculated from the model and the prior is specified as a design input.
The normalisation factor (evidence) is the following integral:
\begin{equation}
	p(\mathbf{x}) = \int_{\bm{\theta}}p(\mathbf{x}|\bm{\theta})p(\bm{\theta})d\bm{\theta}.
	\label{eq:evidence}
\end{equation}
For general multidimensional distributions the integral~\eqref{eq:evidence} becomes intractable.
Consequently, getting the posterior in~\eqref{eq:BT} becomes infeasible, hence the need for the apximate approaches.

\subsection{Markov Chain Monte Carlo}
\gls{mcmc} algorithm is a ubiquitous method for solving integration problems \eqref{eq:evidence} in various areas, e.g. statistics, physics and econometrics.
Central in the approach is the way of taking samples, say a set of $N$ samples $\bm{\theta}^{(i)}\in\mathbb{R}^m,\ i=1,...,N$, from a target density $p(\bm{\theta})$.
Using these $N$ samples, we can approximate $p(\bm{\theta})$,  by calculating the empirical point-mass function
$$p_N(\bm{\theta})=\frac{1}{N} \sum_{i=1}^{N} \delta_{\bm{\theta}^{(i)}}(\bm{\theta}),$$
where $\bm{\theta}^{(i)}$ is the $i^{th}$ sample in our set $\mathbf{\theta}$ and $\delta_{\bm{\theta}^{(i)}}$ is its delta-Dirac mass.

In our case we used \gls{mcmc} with \gls{nuts}~\cite{Hoffman2011}, implemented in Python with the PyMC3~\cite{Salvatier2016} library.
\Gls{nuts} sampling is an extension of the well known \gls{hmc}~\cite{Betancourt2017}.
\gls{hmc} tends to be sensitive to the required user inputs.
This is mostly avoided since the \gls{nuts} algorithm stops automatically when it starts retracing its steps.
Additionally, authors of the \gls{nuts} developed a method for automatic adaptation of the step size, so that the sampler needs minimal user-defined parameters on entry.
For the experimental data described later on, we have let the algorithm run for $300.000$ iterations, which proved sufficient to guarantee convergence.

\subsection{\Glsdesc{vb}}
Using \gls{mcmc} methods for solving~\eqref{eq:evidence} can produce results very close to the true posterior distribution.
However, for the multidimensional cases, the computational load and sheer number of samples required for obtaining proper estimate of the posterior blows up.
One solution to this problem is to find a sufficiently close approximation of the posterior with the significantly lower computational load. 

The main idea of the \gls{vb} approach is finding a candidate distribution $q_{\bm{\lambda}}(\bm{\theta})$ (parameterized with the hyperparameters \footnote{A hyperparameter is a parameter of a prior distribution; the term is used to distinguish them from parameters of the model.} $\bm{\lambda}\in\mathbb{R}^{\nu}$), that is a \textit{sufficiently close approximation} of the \textit{true posterior} $p(\bm{\theta}|\mathbf{x})$.
The distribution $q_{\bm{\lambda}}(\bm{\theta})$ is usually referred to as the \emph{variational distribution}.
The variational distribution is typically selected from the mean-field variational family~\cite{Blei2016}.
This means that one can assume independence among the latent variables of the variational distribution.

\begin{figure}[h]
	\centering
	\includegraphics{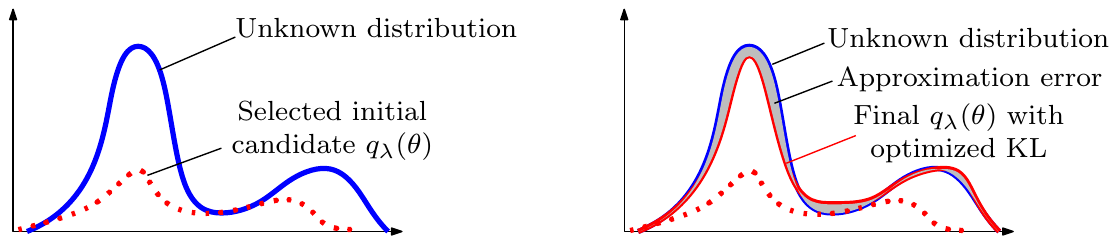}
	\caption{Optimization process of finding the closest variational distribution $q_{\bm{\lambda}}(\bm{\theta})$ over the set of latent variables $\bm{\lambda}$.}
	\label{fig:vi}
\end{figure}

A variational distribution that is the best fit for the true posterior can be obtained by minimizing  the \gls{kl} divergence~\cite{Blei2016}, which can be re-arranged as follows:
\begin{equation}\label{eq:kl}
	\begin{split}
		\gls{kl}(q_{\bm{\lambda}}(\bm{\theta})||p(\bm{\theta}|\mathbf{x})) &=\mathbb{E}_q\left[ \log \frac{q_{\bm{\lambda}}(\bm{\theta})}{p(\bm{\theta}|\mathbf{x})}\right]\\
		& = \mathbb{E}_q [\log q_{\bm{\lambda}}(\bm{\theta}) ] - \mathbb{E}_q [ \log p(\bm{\theta}|\mathbf{x}) ]\\
		& = \mathbb{E}_q [ \log q_{\bm{\lambda}}(\bm{\theta}) ]- \mathbb{E}_q  [\log p(\mathbf{x},\bm{\theta}) - \log p(\mathbf{x}) ]\\
		& = \mathbb{E}_q [\log q_{\bm{\lambda}}(\bm{\theta})] - \mathbb{E}_q [\log p(\bm{\theta},\mathbf{x})] + \log p(\mathbf{x})\\
		& = - ({\mathbb{E}_q[\log p(\bm{\theta},\mathbf{x})] - \mathbb{E}_q [\log q_{\bm{\lambda}}(\bm{\theta})]}
		+ \log p(\mathbf{x})\\
		& = - \underbrace{(\mathbb{E}_q[\log p(\mathbf{x}|\bm{\theta})p(\bm{\theta})] - \mathbb{E}_q [\log q_{\bm{\lambda}}(\bm{\theta})])}_{L(\bm{\lambda},\bm{x})} + \log p(\mathbf{x})
	\end{split}
\end{equation}
Second term $\log p(\mathbf{x})$ is constant.
The first term $L(\bm{\lambda})$is known as \gls{elbo} and by maximising it one can minimize the \gls{kl} divergence between the variational distribution and the true posterior.
So the goal is to solve the following optimization problem:
\begin{equation}
	\bm{\lambda}^*=\underset{\bm{\lambda}\in\Omega}{\arg\min}\gls{kl}(q_{\bm{\lambda}}(\bm{\theta})||p(\bm{\theta}|\mathbf{x})) =  \underset{\bm{\lambda}\in\Omega}{\arg\max} {L(\bm{\lambda},\bm{x})},
\end{equation}
where ${\Omega}$ is the set of all possible values of the hyperparameters $\bm{\lambda}$.
The overall idea is schematically presented in \figurename~\ref{fig:vi}.
It is assumed that the model parameters are random variables.
However, their true distribution is almost always unknown.
In such a case, the best option is to select an approximate candidate distribution that will be used instead.
When selecting the candidate distribution, we try to incorporate prior knowledge as much as possible, in particular the support interval, previous empirical observations, etc. 
Some mismatch between the selected distributions and true posterior is usually always present, which causes a bias in the \gls{vb} solution.
However, this is a small price to pay compared to the substantial increase in the computational efficiency of the \gls{vb} approach.
The \gls{vb} algorithm was implemented in Python using the PyTorch~\cite{NEURIPS2019_9015} library.
For a link to the implementation and an example data set sufficient to recreate a numerical example, see \ref{supp_mat}.

\subsection{Finding the optimal hyper-parameters}
Using \gls{vb} we convert the problem of finding the posterior distributions from  statistical inference to a much simpler task of optimizing a cost function for a variational family parameterised by the vector of hyperparameters $\bm{\lambda}$.
The cost function in this case is \gls{elbo}.
For the optimization part of our algorithm we use the \gls{adam} optimizer~\cite{kingma2014adam}.

\Gls{adam} allows an adaptive correction of the learning rates for each element $\lambda^{(s)}, s=1,...,\nu$ of the vector $\bm{\lambda}$ during the run of the algorithm by calculating the first and second moment of the gradient, denoted as ${m}_t^{(s)}$ and $v_t^{(s)}$ respectively.
This is done using exponentially moving averages computed on the gradient $g_t^{(s)}, \ s=1,...,\nu$
\begin{equation}\label{eq:ADAM_rec}
	\begin{split}
		m_t^{(s)} &= \beta_1 m_{t-1}^{(s)} + (1-\beta_1) g_t^{(s)}\\
		v_t^{(s)} &= \beta_2 v_{t-1}^{(s)} + (1-\beta_2) \left (g_t^{(s)}\right)^2,
	\end{split}
\end{equation}
where $t$ denotes the iteration number and $\beta_{1,2}$ are the parameters of the moving average.
The default values for $\beta$ parameters are 0.9 and 0.999 and the initial values of first and second moment of the gradient are both set to $0$, since it turns out this does not impact the convergence rate significantly.
For a set of measured data (impedance in our case) $\bm{x}$, the gradients can be presented with the following vector
\begin{equation}
	\bm{g}_t=\left( g_t^{(1)}, \ldots , g_t^{(\nu)}\right) ^{T} = \nabla_{\bm{\lambda}_t}  L(\bm{\lambda}_t,\bm{x}).
\end{equation}
The first and the second moment are only estimated with $m^{(s)}$ and $v^{(s)}$, therefore we want them to satisfy the following condition
\begin{equation}
	\begin{split}
		\mathbb{E}\{ m_{t}^{(s)} \} &= \mathbb{E} \left( g_t^{(s)} \right)\\
		\mathbb{E}\{ v_{t}^{(s)} \} &= \mathbb{E} \left( (g^{(s)}_t)^2 \right)
	\end{split}
\end{equation}
Above conditions ensure that we are dealing with unbiased estimates.
First moment at step $t$ in the recursive equation \eqref{eq:ADAM_rec} is
\begin{equation}\label{eq:firstmoment_t}
	m_t^{(s)} = \left( 1- \beta_1 \right) \sum_{i=1}^{t} \beta_1 ^{t-i} {g}_i^{(s)} .
\end{equation}
We can see some bias still occurs in this estimate.
Applying expected value operator to equation \eqref{eq:firstmoment_t} gives us
\begin{equation}
	\begin{split}
		E_{\lambda} \left[ m_t^{(s)} \right] &= \left( 1-\beta_1 \right) \sum_{i=1}^{t} \beta_1^{t-1} E_{\lambda} \left [ {g}_i^{(s)} \right]\\
		&\approx \left( 1- \beta_1 \right) \left( \sum_{i=1}^t \beta_1^{t-1} \right) E_{\lambda} \left[ {g}_t^{(s)} \right]\\
		&=\left( 1-\beta_1^t \right) E_{\bm{\lambda}} \left[ {g}_t^{(s)} \right].
	\end{split}
\end{equation}
Bias correction is done automatically by \gls{adam} during the evaluation of $m$ and $v$ as follows
\begin{equation}
	\begin{split}
		\hat{m}_t^{(s)} &= \frac{m_t^{(s)}}{1-\beta_2^t}\\
		\hat{v}_t^{(s)} &= \frac{v_t^{(s)}}{1-\beta_2^t}.
	\end{split}
\end{equation}
Optimization steps are also adjusted.
For each individual parameter the algorithm updates
\begin{equation}
	\lambda _t ^{(s)} = \lambda _ {t-1} ^{(s)} - \eta \frac{\hat{m}_t^{(s)} }{\sqrt{\hat{v}_t ^{(s)}} + \eta},
\end{equation}
where $\eta=0.001$.

The use of \gls{adam} is widely used across many areas and applications since it delivers efficient and fast results.
It should be noted, however, that as a heuristic optimization algorithm there is no general guarantee for convergence. 
Fortunately, \citet{kingma2014adam} proved that the algorithm converges globally in the convex settings.
This was further refined and proved by \citet{Reddi2019}.
We used the default settings of $\beta_1 = 0.9$ and $\beta_2=0.999$, while setting the learning rate depending on the data set at hand.
\mnewn{Reviewer 2, Comment 2.19}
{
	The number of iterations is at least  $8000$ but no more than $35000$.
	An additional stopping criterion was defined by converged  \gls{elbo} values, i.e. when the relative change of less than $1\%$ is achieved over $1000$ iterations.
	In all of the optimization runs the maximum number of iterations were almost never reached. 
	The average number of iterations was around $13000$.}


\section{Numerical example}
\label{sec:numeric}
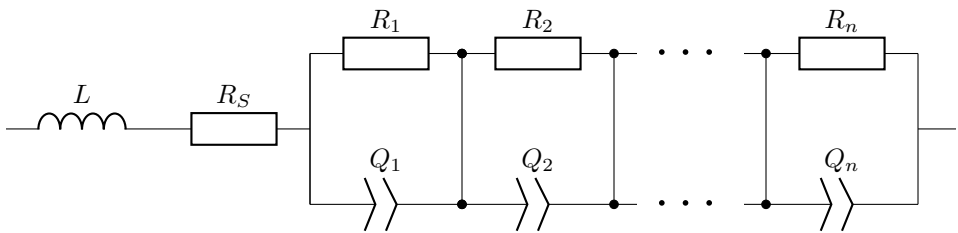
\begin{figure}[h!]
	\centering
	\begin{circuitikz}[scale=1, european]
		\ctikzset{bipoles/empty/voltage/bump b/.initial=20}
		\draw (2,2) to [short] (2,0);
		\draw(-2,1) to [american inductor={$L$}](0,1);
		\draw (0,1) to [resistor={$R_S$}](2,1);
		\begin{scope}[xshift=0cm]
			\draw (2,2) to [short] (2,0);
			\draw (2,2) to [resistor,l=$R_1$,-*](4,2);
			\draw (2,0) to [newcomponent,l=$Q_1$,-*] (4,0);
			\draw (4,2) to [short,*-*] (4,0);
		\end{scope}
		\begin{scope}[xshift=2cm]
			\draw (2,2) to [R,l=$R_2$,-*](4,2);
			\draw (2,0) to [newcomponent,l=$Q_2$,-*] (4,0);
			\draw (4,2) to [short] (4,0);
		\end{scope}
		\begin{scope}[xshift=6cm]
			\draw (0,2) --  (2,2)node[midway,scale=2,fill=white]{$\cdots$};
			\draw (0,0) -- (2,0)node[midway,scale=2,fill=white]{$\cdots$};
		\end{scope}
		\begin{scope}[xshift=6cm]
			\draw (2,2) to [resistor,l=$R_n$,*-](4,2);
			\draw (2,0) to [newcomponent,l=$Q_n$,*-] (4,0);
			\draw (4,2) to [short] (4,0);
			\draw (2,0) to [short] (2,2);
		\end{scope}
		\begin{scope}[xshift=8cm]
			\draw (2,1) -- (2.5,1);
		\end{scope}
	\end{circuitikz}
	\caption{The ECM as a series of  RQ-elements.}
	\label{fig:ecm_general}
\end{figure}

Since we focus on solid oxide fuel cells in this paper, we will first demonstrate the performance of \gls{vb} algorithm using simulated data.
\mnewn{Reviewer 1, Comment 1.2, Reviewer 2, Comment 2.19}{The rationale for the model structure arises from the  electrochemical arguments explained in more detail in section 4. The model can be represented as a serial connection of resistance, inductance and RQ elements
	\begin{equation}\label{eq:3RQ_EIS}
		Z(\omega)=R_s + \sum_{i=1}^N\frac{R_i}{(j\omega)^{\alpha _i}  Q_i  R_i +1} + j\omega L
	\end{equation}
	where $R_s$ stands for serial resistance,  $R_i$ is parallel resistance and $Q_i$ constant-phase  parameter, fractional order of $i^{th}$ pole is denoted with $\alpha_i \in (0,1]$,  $\omega=2\pi f$ and  $f$ is frequency. 
	The schematic of the \gls{ecm} structure is given in \figurename~\ref{fig:ecm_general}. 
	For the simulation study below we take $N=3$.}

\begin{figure}
	\begin{center}
		\begin{tikzpicture}[auto,node distance=2cm]
			\tikzset{
				>=stealth',
				punkt/.style={
					rectangle,
					rounded corners,
					draw=black, thick,
					text width=8.5em,
					minimum height=3.5em,
					node distance=2.0cm,
					text centered,
					execute at begin node=\setlength{\baselineskip}{1.2em}},
				eis/.style={
					text centered,
					execute at begin node=\setlength{\baselineskip}{1.2em}},
				pil/.style={
					->,
					thick,
					shorten <=2pt,
					shorten >=2pt,}
			}
			\node (u) []{$i(t)$};
			\node (H) [process, right of=u]{$H(s)$};
			\node (i) [, right of=H]{$u(t)$};
			\coordinate (CENTER) at ($(u)!0.5!(H)$);
			\node (em1) [circle, draw, black, below of=CENTER]{+};
			\node (em2) [circle, draw, black, right of=i]{+};
			\node (n1) [left of=em1]{$n_1 \sim \mathcal{N}(0,\sigma_1)$};
			\node (n2) [above of=em2]{$n_2 \sim \mathcal{N}(0,\sigma_2)$};
			\node (CWT1) [decision,eis, below of=i,text width=2.5em, align=center]{CWT\\ \cite{Boskoski2017}};
			\node (CWT) [decision,eis, below of=em2,text width=2.5em, align=center]{CWT\\ \cite{Boskoski2017}};
			\node (pos)[] at ($(CWT)!0.5!(CWT1)$) {};
			\node (EIS) [punkt, below of=pos]{Impedance evaluation};
			\node (Z) [,below of=EIS, align=center]{EIS curve as shown in \figurename~\ref{fig:3rqnumerical1000}};
			\draw [arrow] (u) -- (H);
			\draw [arrow] ($ (u) !.5! (H) $) -- (em1);
			\draw [arrow] (H) -- (i);
			\draw [arrow] (i) -- (em2);
			\draw [arrow] (n1) -- (em1);
			\draw [arrow] (n2) -- (em2);
			\draw [arrow] (em2) -- node [midway,text width=2.5em] {measured \\voltage}(CWT);
			\draw [arrow] (em1) -- node [above]{measured}  node [below]{current}  (CWT1);
			\draw [arrow] (CWT) -- (EIS);
			\draw [arrow] (CWT1) -- (EIS);
			\draw [arrow] (EIS) -- (Z);
		\end{tikzpicture}
	\end{center}
	\caption{Simulation of \gls{eis} data with additive noise in input and output.}
	\label{noisesim}
\end{figure}
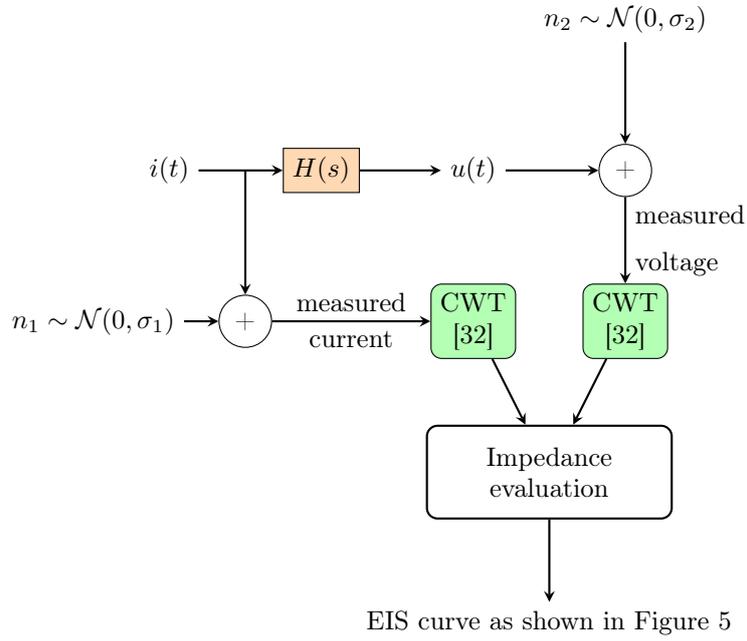

The ``true'' impedance \eqref{eq:3RQ_EIS} was simulated on the frequency interval $f\in[10^{-4}, 10^4]$ Hz by using parameters listed in \tablename~\ref{tb:3rq}.
The additive measurement noise was applied separately to the voltage $u(t)$ and current $i(t)$ as $n_1(t)$ and $n_2(t)$ respectively.
It should be noted that $n_1(t)$ and $n_2(t)$ are zero-mean and uncorrelated.
Having the simulated input and output signals, the transfer function was estimated from Morlet wavelet transform of the input and output as described in~\cite{Boskoski2017}.
The entire process is presented in \figurename~\ref{noisesim}.

\begin{table}[h]
	\centering
	\caption{Model parameters and variational distributions calculated using equations \eqref{eq:lognorm} and \eqref{eq:beta}.}
	\begin{tabular}{lr} 
		\toprule
		
		True parameter & Variational distribution\\
		\midrule
		$R_s$ = $3$ $ \Omega$ & $\sim \text{Lognormal}(0.84,0.39)$ \\
		\midrule
		$R_1=1$ $ \Omega$, $R_2=2$ $ \Omega$, $R_3 =3$ $ \Omega$ & $ \sim \text{Lognormal}(1.59,0.20)$\\
		\midrule
		$Q_1=0.1$ $F s^{\alpha_1}$& $\sim \text{Lognormal}(-0.35,0.83)$ \\
		$Q_2=5$ $F s^{\alpha_2} $&$\sim \text{Lognormal}(1.96,0.83)$ \\
		$Q_3=150$ $F s^{\alpha_3}$&$ \sim \text{Lognormal}(4.99,0.55)$ \\
		\midrule
		$\alpha_1 = 0.88$, $\alpha_2 = 0.82$, $\alpha_3 = 0.99$&$ \sim \text{Beta}(13.91,5.68)$ \\
		\midrule
		$L= 100~nH$\\
		\toprule
	\end{tabular}
	\label{tb:3rq}
\end{table}

\begin{figure}
	\centering
	\includegraphics{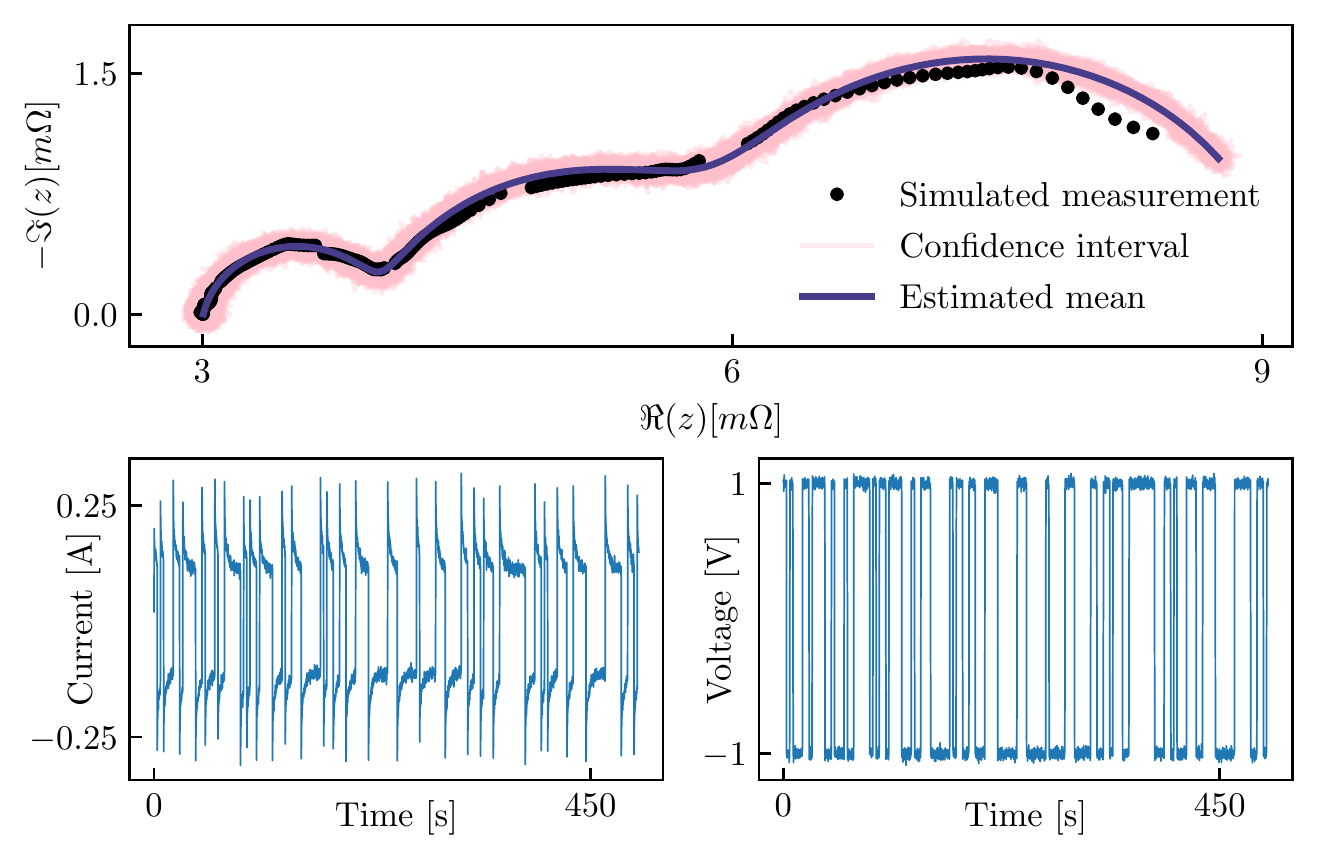}
	\caption{Results of the \gls{vb} algorithm (above). Current and voltage signals used in simulation (below).}
	\label{fig:3rqnumerical1000}
\end{figure}

The first step in applying the \gls{vb} approach is the definition of variational distributions that approximate the true posterior $p(\bm{\theta}|\mathbf{x})$.
They are listed in \tablename~\ref{tb:3rq}.
For our case $\bm{\theta} \in \mathbb{R}_+^{10}$, with additional limitation as $\alpha_i\in[0,1]$ .
Beta distribution was therefore chosen as the best fit for the $\alpha_i, i \in \{ 1,2,3 \} $ parameters and Log-normal distribution for the rest.
We selected the means and variances for our variational distributions and then determined their parameters using \eqref{eq:lognorm} and \eqref{eq:beta}.

Rough estimates of  variational distributions can be assessed  from the \gls{eis} curve in \figurename~\ref{fig:3rqnumerical1000}, previous experience and integrating experts knowledge.
In the upper plot in \figurename~\ref{fig:3rqnumerical1000} one can notice that the real part of the curve starts around $3$ $\Omega$.
This provides a rough estimate for the parameter $R_s$. Hence  selecting variational distribution with mean $2.5$~$\Omega$ and variance to $1$ seems a reasonable choice.
The next step is setting up the variational distributions for the parameters $R_{1,2,3}$.
Just by looking at the axis values, it is apparent that they should be located within the interval (1,10)~$\Omega$.
Mean values of their distributions were therefore set in the middle of the interval.
Variational distributions of the parameters $Q_{1,2,3}$ were roughly estimated with the help of experiences gained from previous experiments and by incorporating experts knowledge.
Lastly, the fractional order powers $\alpha_{1,2,3}$ are expected to have values in the interval [0.5,1] so we set their variational distributions with means at $0.75$.

Once we have determined the mean and variance of our variational distributions, we have to transform them into the correct parameters for the chosen distributions.
To obtain the parameters of the log-normal distribution $(\sigma _{ln} , \mu _{ln})$ from the mean and variance $(\sigma,\mu)$, the following equations can be used:
\begin{equation}\label{eq:lognorm}
	\sigma _{ln} = \ln \left( \frac{\sigma ^2}{\sqrt{\sigma ^2 + \mu ^2}} \right) \hspace{2cm} \mu _{ln} = \ln \left( 1 + \frac{\mu ^2}{\sigma ^2} \right)
\end{equation}
and for beta distribution parameters $\alpha$ and $\beta$ e.g. $\text{Beta}(\alpha,\beta)$ we use:
\begin{equation}\label{eq:beta}
	\alpha = \frac{\sigma ^2 - \sigma ^3}{\mu} - \sigma \hspace{2cm} \beta = \frac{\alpha}{\sigma} - \alpha
\end{equation}

\begin{figure}[htbp]
	\centering
	\includegraphics{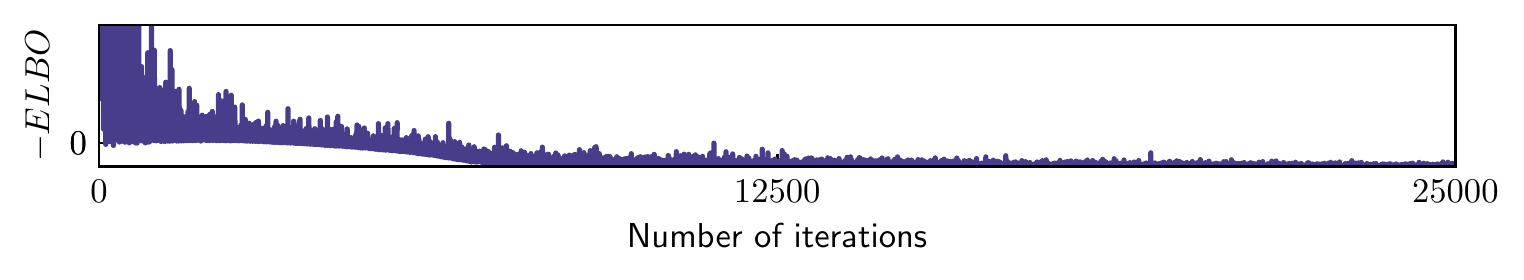}
	\caption{The evolution of ELBO loss 
		during  optimization process.}
	\label{fig:3rqelbo}
\end{figure}

The optimization was performed using a global learning rate of $0.05$ and was completed after $25000$ steps.
\Gls{elbo} loss curve can be observed in \figurename~\ref{fig:3rqelbo}.
By sampling the resulting posterior distributions and simulating the model~\eqref{eq:3RQ_EIS} it is possible to visualize the obtained results.
For this purpose, $1000$ samples were drawn for each parameter from their respective posterior distribution.
Simulating the $1000$ \gls{eis} curves with the sampled sets of parameters gave us an estimate of the confidence interval of our results, which can be found in \figurename~\ref{fig:3rqnumerical1000}.
The \gls{eis} curve obtained from the means of the posterior distributions for the parameters can also be found on the same \figurename.
The variance of the posterior distributions is effectively presented with the sampled curves and it is bound by the noise variance parameter.
The mean \gls{eis} curve seems to be a sufficiently good fit for the input data.

\mnewn{Reviewer 2, Comment 2.4}{Additional results demonstrating the performance of the  \gls{vb} algorithm on simulated data with heavy additive noise  can be found in \ref{Additional_numerical}.}

\begin{figure}[htbp]
	\centering
	\includegraphics{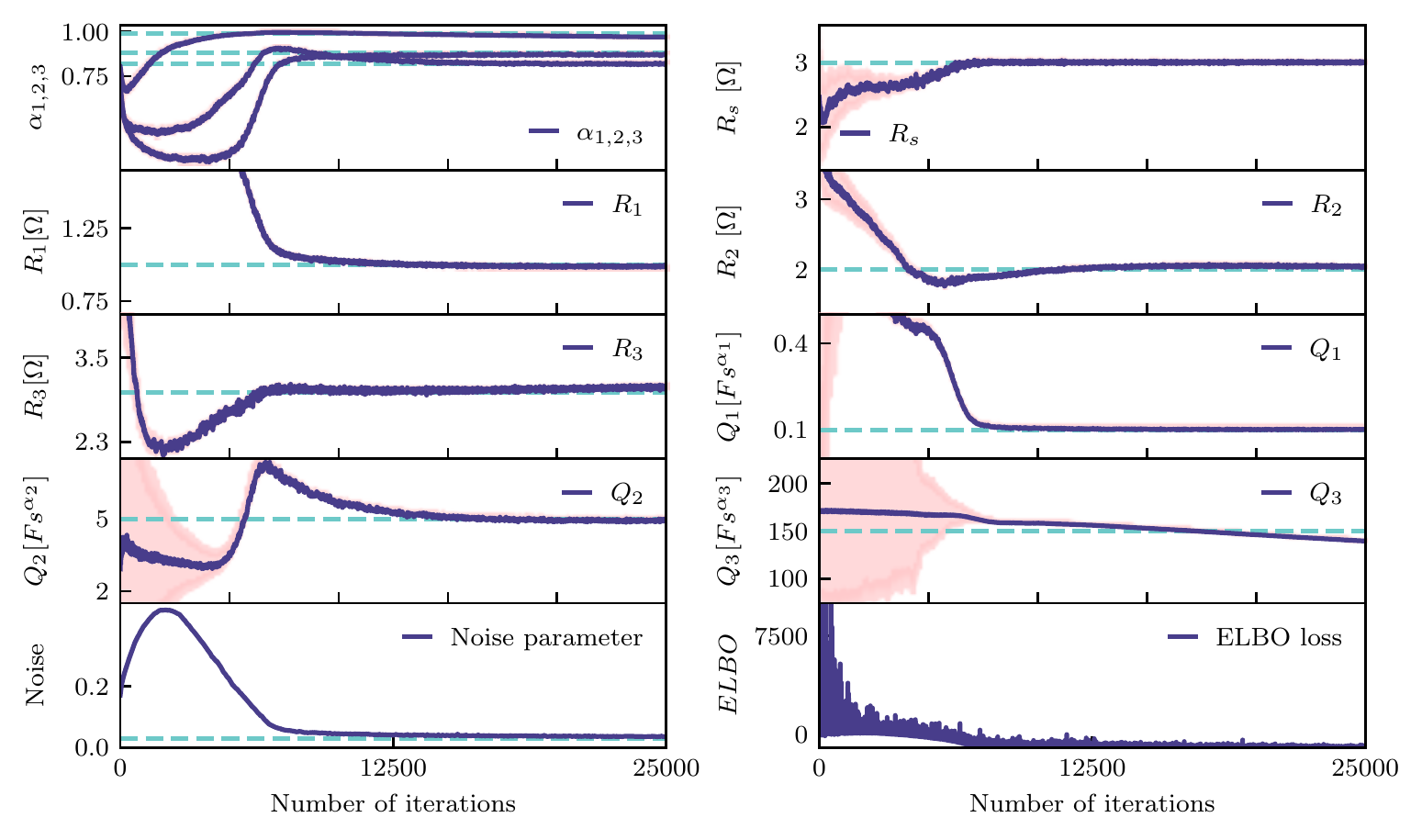}
	\caption{Progress of parameter estimation during the optimization process, true values of parameters used for simulation are represented with dashed lines. The variance of the estimated distribution(pink) is quite low.
		The selected optimal shape of the posterior distributions are shown in \figurename~\ref{fig:alldist}. }
	\label{fig:3rqparameterprogression}
\end{figure}

The evolution of the optimization process for each parameter is presented in \figurename~\ref{fig:3rqparameterprogression}, where we can also find the true parameters used for the simulation of our input data.
We can see that each parameter reaches its true value, except $Q_3$.
We can assume this is the effect of noise, which seems to be greater in the third arc, as seen in \figurename~\ref{fig:3rqnumerical1000}.
Progress in confidence of the algorithm can be easily observed with the help of the evolution of estimated spread in parameter distributions.
As seen in the \figurename~\ref{fig:3rqparameterprogression}, the spread is wide at the beginning and as the optimization progresses and explores larger search space it steadily narrows.

\begin{figure}[htp]
	\centering
	\includegraphics{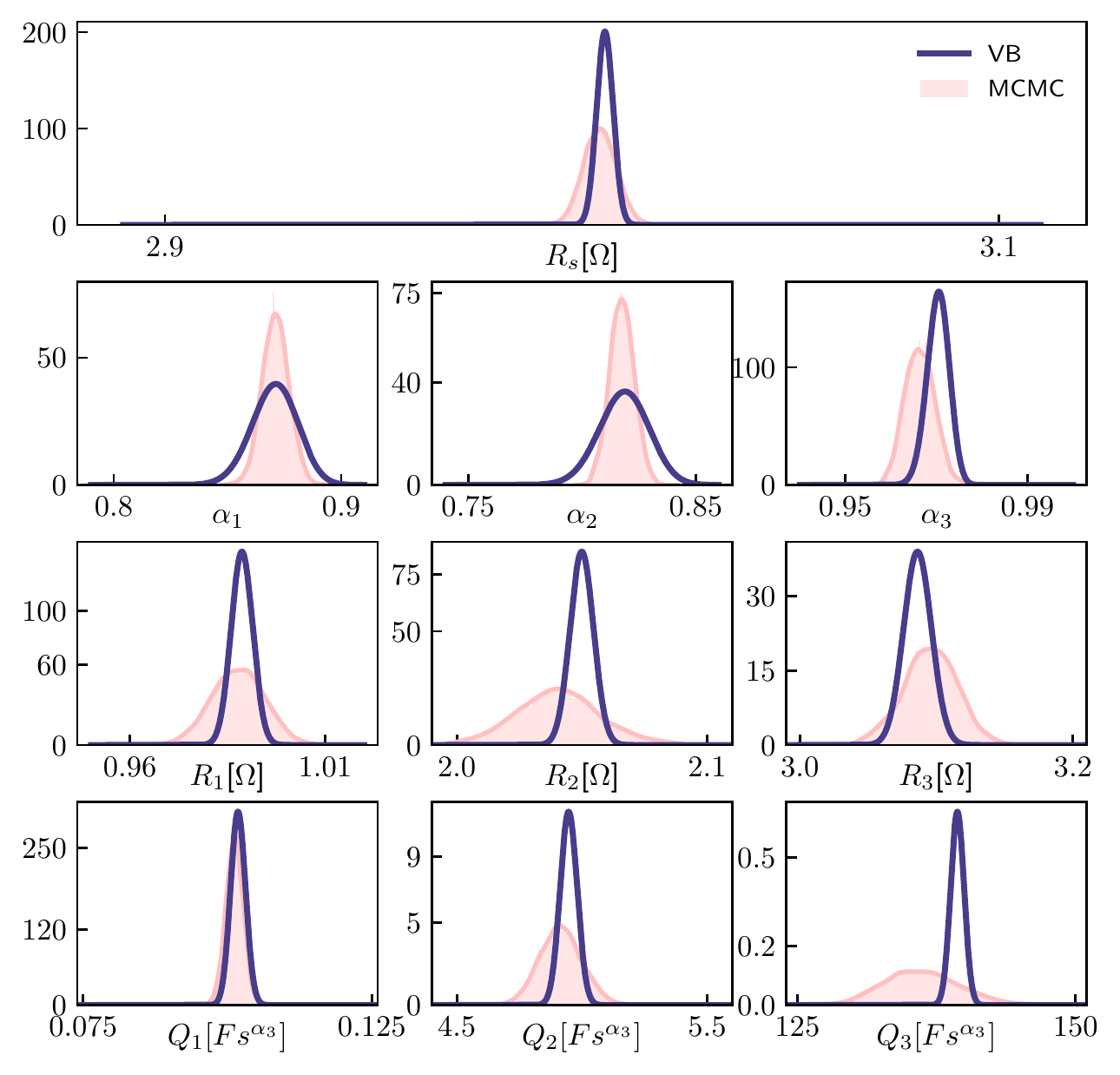}
	\caption{Posterior distributions of parameters derived from simulated data.}
	\label{fig:alldist}
\end{figure}

The resulting posterior distributions of the model parameters as well as the posteriors obtained with \gls{mcmc} are shown in \figurename~\ref{fig:alldist}.
Three key observations can be made.
First, comparing posterior and variational distributions, the optimization process was capable of reaching the ``true'' mean values even for cases when the variational means were ``far'' from the simulated values.
Second, the scales (variance) of the posterior distributions are low, leading to the conclusion that the obtained model has low uncertainty.
\tablename~\ref{tb:post_3rq} lists all the resulting posterior parameter distributions.
Third, the empirical posteriors obtained through \gls{mcmc} method are similar to the ones obtained by \gls{vb} approach.
It should be noted that in some cases \gls{vb}  posteriors can be under-dispersed (lower variance) compared to the ones obtained by \gls{mcmc}.
Being overconfident is a well known behavioural trait of \gls{vb} algorithm and a mathematical proof of this behaviour under certain conditions is provided by Blei and Wang~\cite{wang2019variational,Wang_2018}.
Our results seem to fit the \gls{mcmc} results reasonably well, with  minor overconfidence in the posterior distributions of $R_{1,2}$ and $Q_3$.

\begin{table}[h]
	\centering
	\caption{Posterior distributions of parameters.}
	\begin{tabular}{ccc} 
		\toprule[1pt]
		\multicolumn{3}{c}{		$R_s\sim \text{Lognormal}(1.0,0.0006)$} \\
		\midrule
		$R_1\sim \text{Lognormal} (-0.011,0.003)$ &
		$R_2\sim \text{Lognormal} (0.72,0.002)$  &
		$R_3\sim \text{Lognormal} (1.12,0.003)$
		\\
		
		\midrule
		$Q_1\sim \text{Lognormal} (-2.28,0.0007)$ &
		$Q_2\sim \text{Lognormal} (1.6,0.007)$  &
		$Q_3\sim \text{Lognormal} (4.94,0.004)$
		\\
		\midrule
		$\alpha _1\sim \text{Beta} (969.35,144.11)$ &
		$\alpha _2\sim \text{Beta} (1016.18,225.57)$  &
		$\alpha _3\sim \text{Beta} (4724.72,144.39)$
		\\
		
		\bottomrule[1.5pt]
	\end{tabular}
	\label{tb:post_3rq}
\end{table}

\section{Experimental validation}
\label{sec:results}
The \gls{vb} was validated using \gls{eis} curves measured on stack of $6$ anode supported solid oxide fuel cells which were installed in an insulated ceramic housing.
\mnew{Reviewer 2, Comment 2.4}{Stack operated for a total of 4500 hours at the operating temperature of $750$ °C.
The active area of a single cell was 80 cm$^2$.
The \gls{sofc} short stack was fed with a mixture of hydrogen and nitrogen with a flow rate \ce{H2}/\ce{N2}=0.216/0.144 Nl h$^{-1}$cm$^{-2}$ whereas the air flow rate was 4 Nl h$^{-1}$cm$^{-2}$.
Stack was operated at nominal current of 32~\si{\ampere} (0.4~\si{\ampere\per\square\cm}) with fuel utilization of 77.5 \%.

The \gls{eis} curves were obtained by current excitation (galvanostatic mode) with discrete random binary signal.
The excitation current had DC value of I${_\text{DC}}$=32 A with peak-to-peak amplitude of 2 A. 
The amplitude was chosen low to ensure linearised stack response, but still large enough to guarantee  sufficient signal-to-noise ratio.

The current and voltage sensor have sufficiently wide frequency bandwidth with cut-off frequency of 240~kHz.
The cell voltages were measured independently using a differential 16-bit NI USB-6215 data acquisition system.
The analogue signals were firstly low pass filtered at 10.8~kHz and sampled with sampling frequency $f_s=50$~kHz. 
Additional information about the \gls{sofc} short stack in question can be found in \cite{NUSEV2021229491}.}

\subsection{Experimental results}
\begin{figure}[htp]
	\centering
	\includegraphics{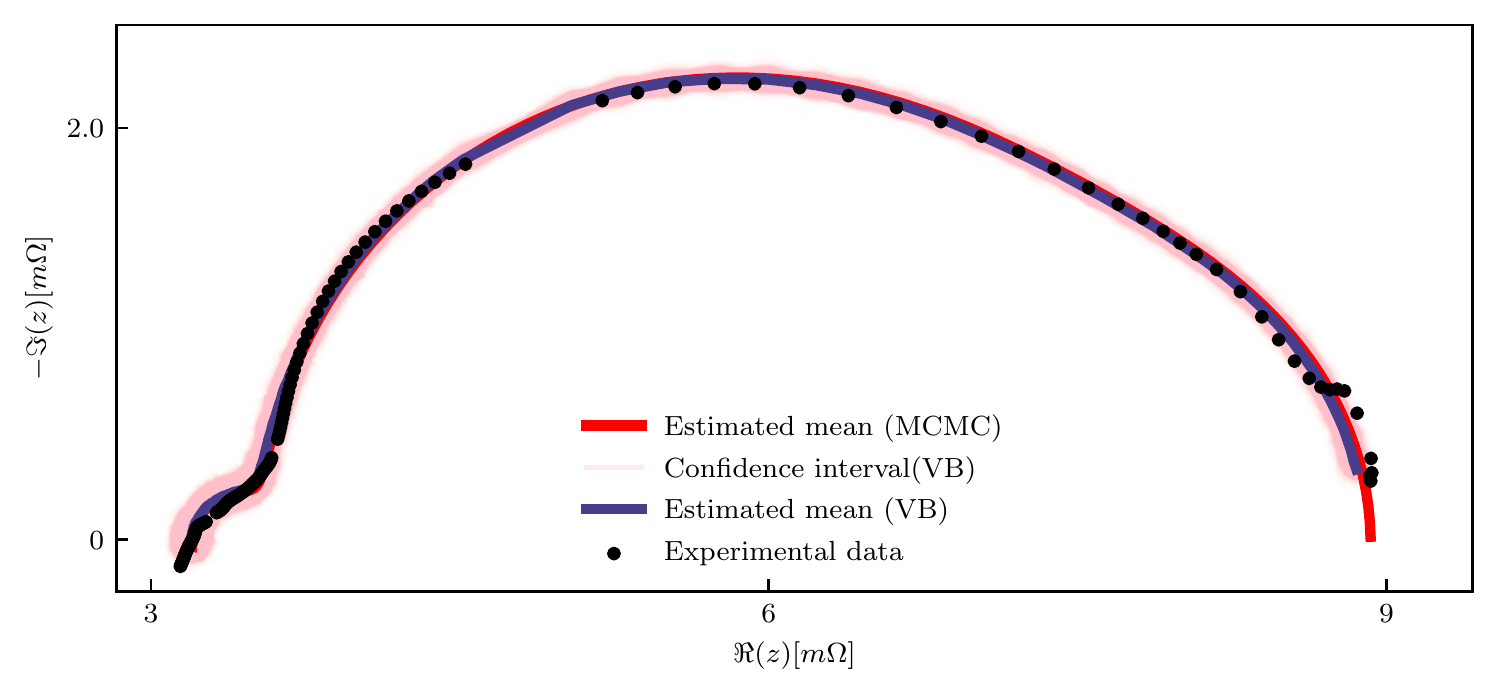}
	\caption{Measured \gls{eis} and the estimated mean \gls{eis} values from \gls{mcmc} and \gls{vb}.}
	\label{fig:exp1000sampled}
\end{figure}

The measured \gls{eis} curve is shown in \figurename~\ref{fig:exp1000sampled}.
Note here that the noise in measurements is quite low thanks to the high-quality data acquisition equipment and the well controlled laboratory conditions for the experiment.

\begin{figure}[!htbp]
	\centering
	\includegraphics{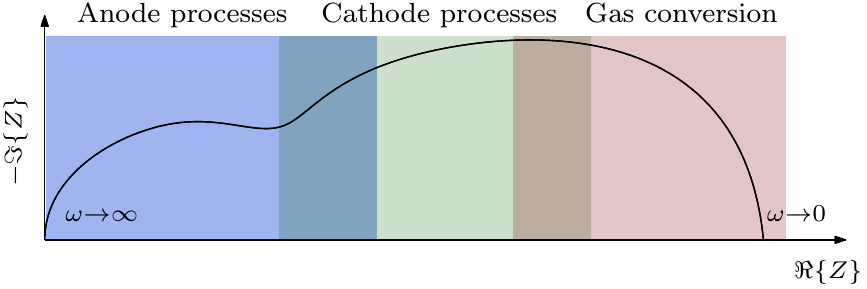}
	\caption{Main groups of processes within the hydrogen-powered \gls{sofc}.}
	\label{fig:eisfreqintervals}
\end{figure}

\mnewn{Reviewer 1, Comment 1.2}{
	The first step in model identification is the selection of the model structure. In our case that breaks down to the selection of $N$, the number of RQ elements and to do so physical arguments are used.
	In case of hydrogen as a fuel one could expect five dominant processes divided into three main groups~\cite{LANG20087509}: gas conversion, cathode processes and anode processes.
	Their time constants span the frequency band from 0.1 Hz to 1 Mhz.
	The gas conversion processes are related to the low frequency part of the \gls{eis} curve, the cathode processes are mainly visible in the mid-frequency range and the high-frequency part can be attributed to the anode related processes.
	The contribution of each group can be seen in \figurename~\ref{fig:eisfreqintervals}.}

Variational distributions were selected by examining the measured \gls{eis} curve.
If we look at the \figurename~\ref{fig:exp1000sampled}, we can see that the high frequency part of the real component has values around $3\text{m}\Omega$, which will serve as the mean value for the $R_s$ parameter.
The scale of its distribution was set quite large to allow the optimization process to explore broad intervals in the search for the optimum.
Variational distributions of $R_{1,2,3}$ parameters can be inferred from the complex and real values our \gls{eis} curve takes that their values should be roughly between the interval [0.1,3]~m$\Omega$.
From previous testing and by incorporating experts knowledge we can assume that the $Q_{1,2,3}$ parameters will be at least by an order of $10$ apart, so we assumed they are located at $1, 50$ and $500$ $ F s^{\alpha_1}$ respectively.
The scales for their variational distributions were set respectively large.
Finally, for $\alpha_{1,2,3}$ we can assume from experience that they probably take values between $0.5$ and $1.0$, therefore the means of their variational distributions were set at $0.75$. 
All the variational distributions can be seen on \tablename~\ref{tb:exp}.

\begin{table}[h]
	\centering
	\caption{Variational distributions of parameters used for estimation of parameters for experimental data.}
	\begin{tabular}{ccc} 
		\toprule[1pt]
		$R_s\sim \text{Lognormal}(-5.86,0.32)$ &
		$R_{1,2,3}\sim \text{Lognormal} (-8.41,1.27)$ &
		$\alpha_{1,2,3}\sim \text{Beta} (12,3)$\\
		
		\midrule
		$Q_1\sim \text{Lognormal} (-2.31,2.14)$ &
		$Q_2\sim \text{Lognormal} (3.57,0.83)$  &
		$Q_3\sim \text{Lognormal} (6.20,0.20)$
		\\
		
		\bottomrule[1.5pt]
	\end{tabular}
	\label{tb:exp}
\end{table}

The learning rate was set to $0.005$.
This value is lower than in the numerical example, since the goal was to allow steady and efficient estimation.
The optimization finished after $35000$ steps.

The performance of the algorithm can best be observed by looking at the \gls{eis} curves obtained from the posterior distributions of the estimated parameters.
In the same manner as with numerical example, we also took $1000$ samples from each of parameters posterior distribution, to simulate $1000$ \gls{eis} curves.
The confidence interval is shown in \figurename~\ref{fig:exp1000sampled}, together with the \gls{eis} curve obtained from the posterior means.
The results confirm that despite the low number of iterations, the resulting parameters represent an accurate fit for the measured \gls{eis} data.

\begin{figure}[htp!]
	\centering
	\includegraphics{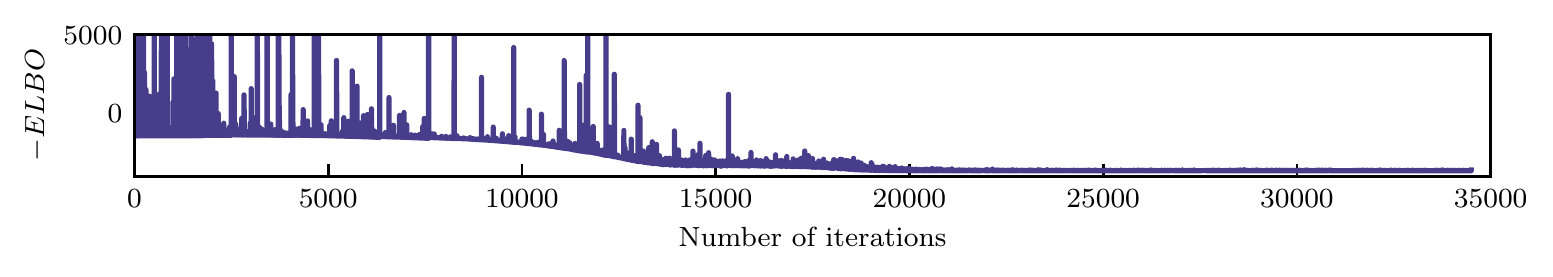}
	\caption{The evolution of ELBO for experimental data.}
	\label{fig:elboexp}
\end{figure}

The convergence of \gls{elbo} can be seen in \figurename~\ref{fig:elboexp}.
Comparing with the \figurename~\ref{fig:3rqelbo} we can notice that the rate of convergence is slower, which can be associated with the much smaller learning rate and smaller number of iterations used for the run on experimental data.
As optimization  approaches the end, the \gls{elbo} value changes minimally, which means the optimal estimates have been found. 

The evolution of parameter estimation is shown in \figurename~\ref{fig:paramprogexp}.
The algorithm steadily converges towards the optimum, while the estimated spread for each parameter continues to decrease.
The parameter evolution is very smooth for each parameter.
Posterior distributions are shown in \figurename~\ref{fig:alldistexp}, where we compare them with results of \gls{mcmc} algorithm. 
We can see the expected slight overconfidence of \gls{vb}, while the results are still inline with \gls{mcmc} results.
It is apparent that scales (variance) of the posterior distributions are quite low.
Posterior parameter distributions are listed in \tablename~\ref{tb:post_exp}. 

\mnewn{Reviewer 1, Comment 1.3;Reviewer 2, Comment 2.4}{Additional \gls{vb} approach results on different settings for \gls{sofc} can be found in \ref{Additional_experimental}, including a result on a measurement recorded after a leakage fault has occured.}

\begin{table}[h]
	\centering
	\caption{Posterior distributions of the parameters.}
	\begin{tabular}{ccc} 
		\toprule[1pt]
		\multicolumn{3}{c}{		$R_s\sim \text{Lognormal}(-5.75,8.3e^{-4})$} \\
		\midrule
		$R_1\sim \text{Lognormal} (-7.995,0.011)$ &
		$R_2\sim \text{Lognormal} (-6.33,0.0039)$  &
		$R_3\sim \text{Lognormal} (-5.61,0.0075)$
		\\
		
		\midrule
		$Q_1\sim \text{Lognormal} (2.25,0.0086)$ &
		$Q_2\sim \text{Lognormal} (6.22,0.0025)$  &
		$Q_3\sim \text{Lognormal} (3.96,0.0028)$
		\\
		\midrule
		$\alpha _1\sim \text{Beta} (1081.56,41.42)$ &
		$\alpha _2\sim \text{Beta} (2216.36,187.46)$  &
		$\alpha _3\sim \text{Beta} (399.28,0.495)$
		\\
		
		\bottomrule[1.5pt]
	\end{tabular}
	\label{tb:post_exp}
\end{table}

\subsection{Comparison of Variational Bayes with averaging spectra}
\begin{figure}[!htpb]
	\centering
	\includegraphics{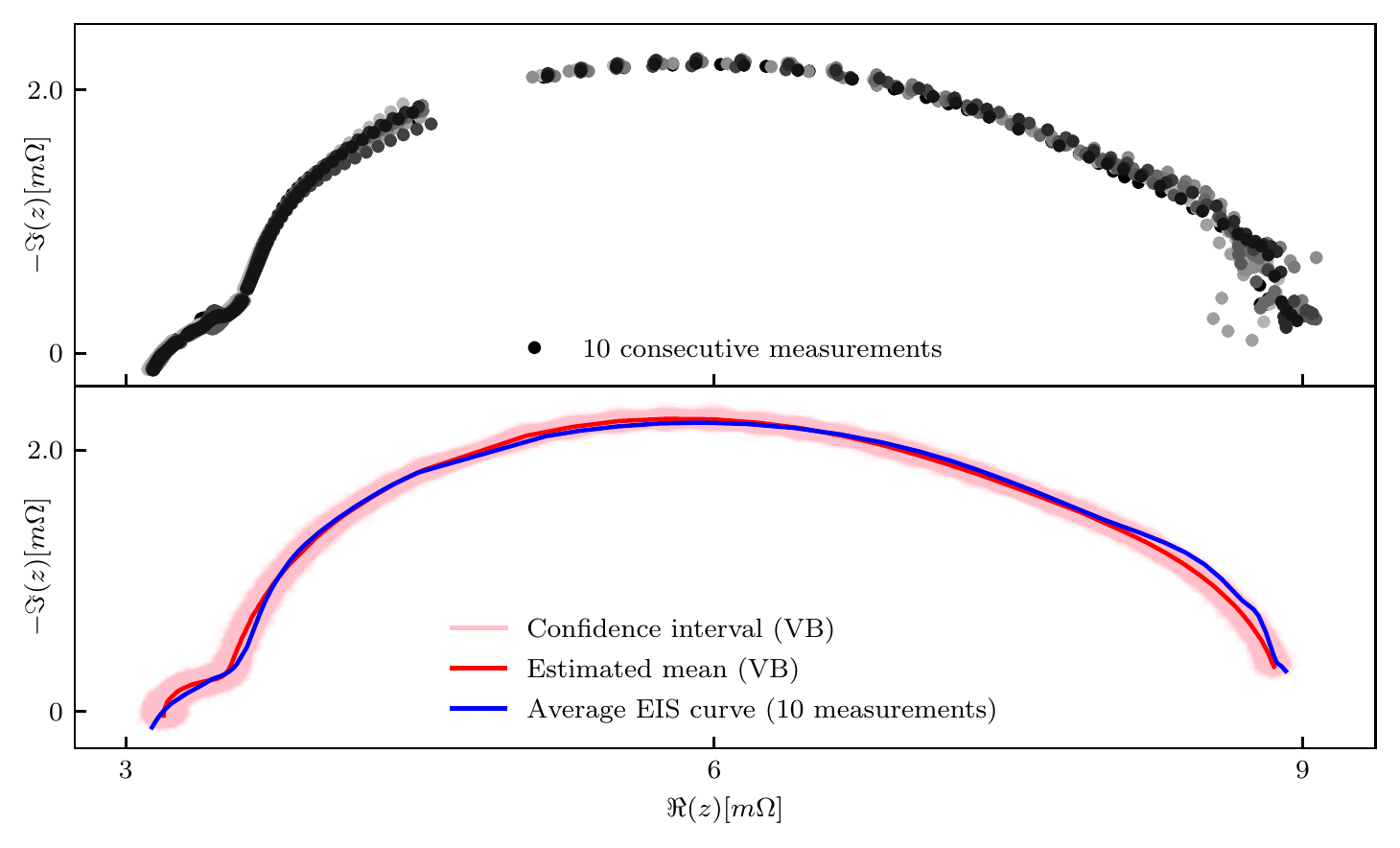}
	\caption{Comparison of \gls{vb} approach and averaging \gls{eis} spectra. Ten  consecutive \gls{eis} evaluations are used (upper part). The compared \gls{vb} posterior and averaged spectra are on the lower graph.}
	\label{fig:averagingandvbmean}
\end{figure}

\mnewn{Reviewer 2, Comment 2.11}{
	Averaging the \gls{eis} obtained from successive  measurements is a pragmatic way to reduce the effects of measurement noise.
	The more spectra are averaged, the better the filtration of noise is. 
	However, in practice that means many repeated perturbations are needed.
	
	Using \gls{vb} approach, we can compute a comparable result by using only one  measurement.
	For the comparison, we averaged 10 consecutive measured spectra and compared with the results of the \gls{vb} approach on the last among them.
	
	The results are presented in \figurename~\ref{fig:averagingandvbmean}, where  measured \gls{eis} data are in the upper part, while the mean estimate from \gls{vb} approach and the evaluated average are presented in the lower part.
	The results of both methods are very similar, even though the \gls{vb} approach  used information from only one measurement. That is practical benefit in the context  of online health monitoring.}

\subsection{Discussion }
Results in \figurename~\ref{fig:exp1000sampled} that the \gls{ecm} identified through the \gls{vb} approach provide an accurate fit for the measured data. 

\begin{figure}[!htbp]
	\centering
	\includegraphics{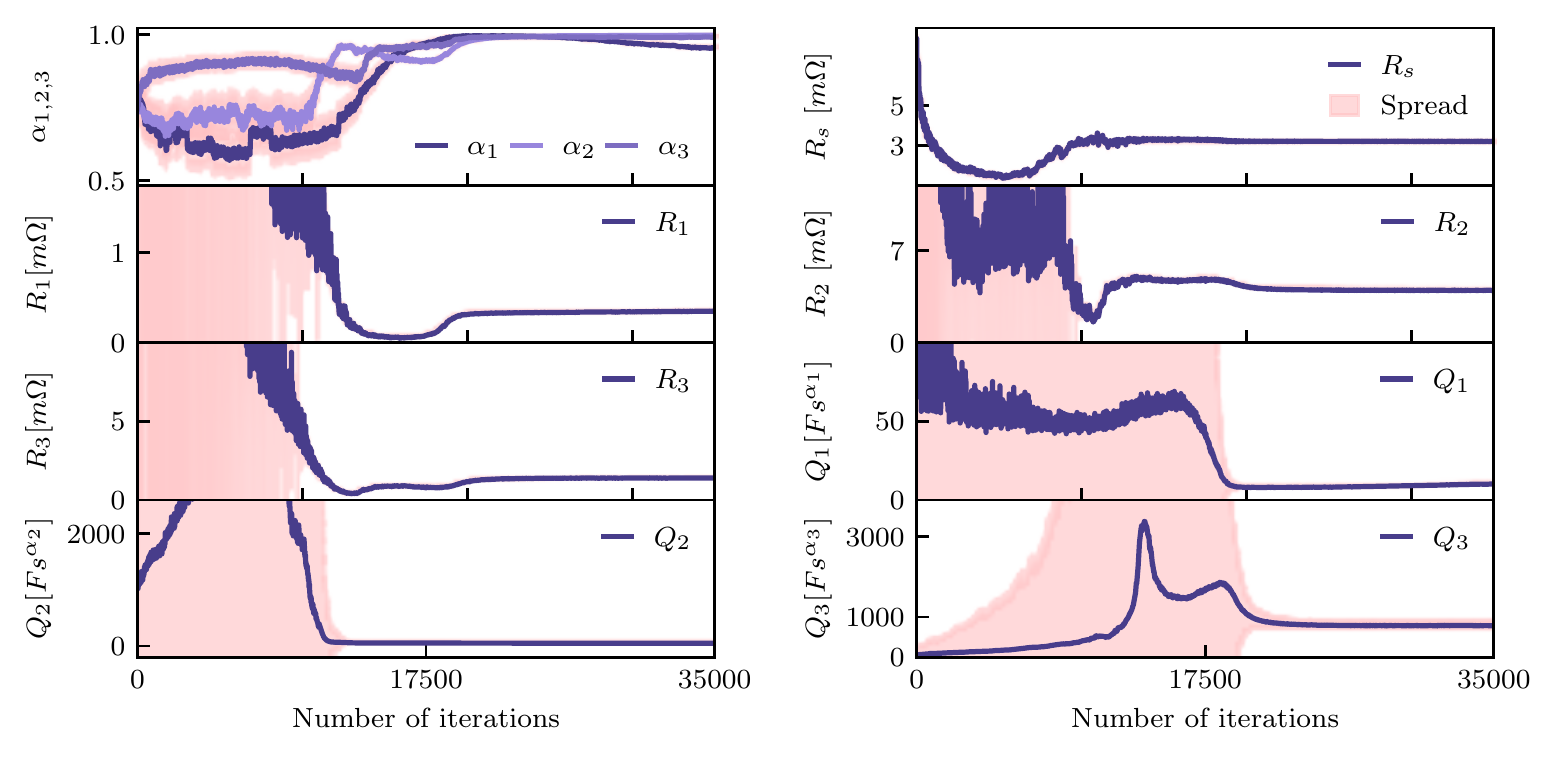}
	\caption{The evolution of the parameter estimates for experimental data. }
	\label{fig:paramprogexp}
\end{figure}

From \gls{elbo} plot in \figurename~\ref{fig:elboexp} it is evident the algorithm converges.
The same observation holds true for the estimated parameters, as shown in \figurename~\ref{fig:paramprogexp}.

\begin{figure}[!htbp]
	\centering
	\includegraphics{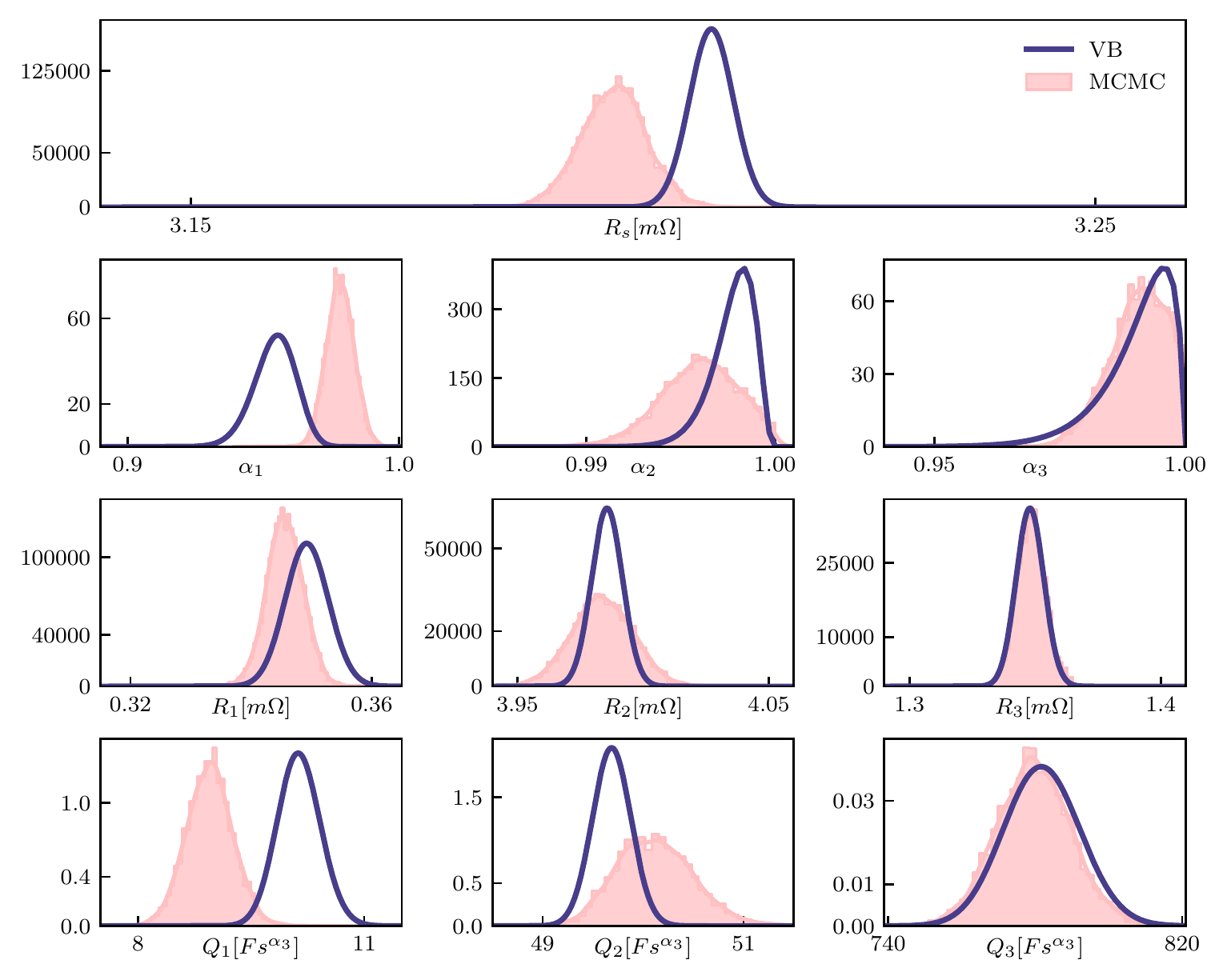}
	\caption{Posterior distributions of parameters inferred from experimental data by  \gls{vb} and \gls{mcmc} algorithm.}
	\label{fig:alldistexp}
\end{figure}

Comparing the resulting posterior distributions from both \gls{vb} and \gls{mcmc} in \figurename~\ref{fig:alldistexp} we can conclude the following.
First, results show  close match for every parameter, with only a slight mismatch noticeable in $R_s, \alpha_1$ and $Q_1$.
Second, the posterior distributions obtained with \gls{vb} have lower dispersion (variance) than the ones obtained through \gls{mcmc}.
As mentioned above, this is common for \gls{vb} approach.

One of the reasons for having overconfident results might be in the selection of the variational distributions.
A possible improvement might be achieved by using more elaborated approximations in terms of a mixture of bounded  distributions~\cite{tenreiro2013boundary}.
That is a topic of further research.

\section{Conclusion}
\mnewn{}{
	In this paper  a novel approach to the statistical estimation of the ECM parameters of \gls{sofc} impedance spectra is presented. 
	From a single \gls{eis} evaluation, the approach is able to evaluate the uncertainty in the model parameters caused by noise and disturbances in the system. 
	The \gls{vb} approach was validated on simulated data as well as on  experimental measurements.
	
	The simulation reveals that even in the case of excessive degree of noise the \gls{vb} algorithm still produces accurate estimates.
	Main benefit of using the method is low computational load when compared to the commonly used \gls{mcmc} approach.
	In particular, to obtain comparable results the time needed was approximately 5 minutes for \gls{vb} approach and around 25 hours for \gls{mcmc} on the same computational power.
	Validation tests indicate that the difference in performance between two methods is minimal, which renders \gls{vb} as an option for on-line \gls{sofc} monitoring. 
	
	Full information on the estimated value and accompanying uncertainty helps avoid misinterpretation of the \gls{eis} data corrupted by noise and disturbances. 
	It is shown that the results of \gls{vb} might be comparable with the results obtained by averaging the spectra. 
	The key advantage of \gls{vb} is that only one evaluated \gls{eis} is needed to get the information for which several spectra in the averaging approach are needed (more spectra means more stack perturbation).
	
	The future work will concentrate on the design of the detection approach that fully exploits the information provided by the \gls{vb}. 
	It is expected that evaluated uncertainties in the estimated parameters will result in a more cautious diagnosis, which will be insensitive to the illogical shapes of the impedance characteristics.
}

	\section*{Acknowledgements}
	The authors acknowledge the support from the Slovenian Research Agency through the programme P2-0001 and the project NC-0003. 
	Part of the support is received through the project RUBY (grant agreement No. 875047) within the framework of the Fuel Cells and Hydrogen
	2 Joint Undertaking under the European Union’s Horizon 2020 research and innovation programme,
	Hydrogen Europe and Hydrogen Europe research. We are grateful to the anonymous referees for their valuable comments and suggestions.
	
	\bibliographystyle{elsarticle-num-names}
	\bibliography{bibstep}

\begin{thebibliography}{35}
\expandafter\ifx\csname natexlab\endcsname\relax\def\natexlab#1{#1}\fi
\providecommand{\url}[1]{\texttt{#1}}
\providecommand{\href}[2]{#2}
\providecommand{\path}[1]{#1}
\providecommand{\DOIprefix}{doi:}
\providecommand{\ArXivprefix}{arXiv:}
\providecommand{\URLprefix}{URL: }
\providecommand{\Pubmedprefix}{pmid:}
\providecommand{\doi}[1]{\href{http://dx.doi.org/#1}{\path{#1}}}
\providecommand{\Pubmed}[1]{\href{pmid:#1}{\path{#1}}}
\providecommand{\bibinfo}[2]{#2}
\ifx\xfnm\relax \def\xfnm[#1]{\unskip,\space#1}\fi
\bibitem[{Lasia(2014)}]{EIS2014}
\bibinfo{author}{A.~Lasia}, \bibinfo{title}{Electrochemical Impedance
  Spectroscopy and its Applications}, \bibinfo{publisher}{Springer-Verlag},
  \bibinfo{address}{New York}, \bibinfo{year}{2014}.
\bibitem[{Nusev et~al.(2021)Nusev, Morel, Mougin, Juričić, and
  Boškoski}]{NUSEV2021229491}
\bibinfo{author}{G.~Nusev}, \bibinfo{author}{B.~Morel},
  \bibinfo{author}{J.~Mougin}, \bibinfo{author}{{\DJ}.~Juričić},
  \bibinfo{author}{P.~Boškoski},
\newblock \bibinfo{title}{Condition monitoring of solid oxide fuel cells by
  fast electrochemical impedance spectroscopy: A case example of detecting
  deficiencies in fuel supply},
\newblock \bibinfo{journal}{Journal of Power Sources} \bibinfo{volume}{489}
  (\bibinfo{year}{2021}) \bibinfo{pages}{229491}. \URLprefix
  \url{https://www.sciencedirect.com/science/article/pii/S0378775321000409}.
  \DOIprefix\doi{https://doi.org/10.1016/j.jpowsour.2021.229491}.
\bibitem[{Mougin et~al.(2019)Mougin, Morel, Ploner, Caliandro, Van~Herle,
  Boškoski, Dolenc, Gallo, Polverino, Pohjoranta, Nieminen, Pofahl, Ouweltjes,
  Diethelm, Leonardi, Galiano, and Tanzi}]{Mougin2019}
\bibinfo{author}{J.~Mougin}, \bibinfo{author}{B.~Morel},
  \bibinfo{author}{A.~Ploner}, \bibinfo{author}{P.~Caliandro},
  \bibinfo{author}{J.~Van~Herle}, \bibinfo{author}{P.~Boškoski},
  \bibinfo{author}{B.~Dolenc}, \bibinfo{author}{M.~Gallo},
  \bibinfo{author}{P.~Polverino}, \bibinfo{author}{A.~Pohjoranta},
  \bibinfo{author}{A.~Nieminen}, \bibinfo{author}{S.~Pofahl},
  \bibinfo{author}{J.~Ouweltjes}, \bibinfo{author}{S.~Diethelm},
  \bibinfo{author}{A.~Leonardi}, \bibinfo{author}{F.~Galiano},
  \bibinfo{author}{C.~Tanzi},
\newblock \bibinfo{title}{Monitoring and diagnostics of sofc stacks and
  systems},
\newblock \bibinfo{journal}{ECS Transactions} \bibinfo{volume}{91}
  (\bibinfo{year}{2019}) \bibinfo{pages}{731--743}.
  \DOIprefix\doi{10.1149/09101.0731ecst}.
\bibitem[{Gallo et~al.(2020)Gallo, Polverino, Mougin, Morel, and
  Pianese}]{GALLO2020115718}
\bibinfo{author}{M.~Gallo}, \bibinfo{author}{P.~Polverino},
  \bibinfo{author}{J.~Mougin}, \bibinfo{author}{B.~Morel},
  \bibinfo{author}{C.~Pianese},
\newblock \bibinfo{title}{Coupling electrochemical impedance spectroscopy and
  model-based aging estimation for solid oxide fuel cell stacks lifetime
  prediction},
\newblock \bibinfo{journal}{Applied Energy} \bibinfo{volume}{279}
  (\bibinfo{year}{2020}) \bibinfo{pages}{115718}. \URLprefix
  \url{http://www.sciencedirect.com/science/article/pii/S0306261920312113}.
  \DOIprefix\doi{https://doi.org/10.1016/j.apenergy.2020.115718}.
\bibitem[{Rakar et~al.(1999)Rakar, Juričić, and Ballé}]{RAKAR1999555}
\bibinfo{author}{A.~Rakar}, \bibinfo{author}{{\DJ}.~Juričić},
  \bibinfo{author}{P.~Ballé},
\newblock \bibinfo{title}{Transferable belief model in fault diagnosis},
\newblock \bibinfo{journal}{Engineering Applications of Artificial
  Intelligence} \bibinfo{volume}{12} (\bibinfo{year}{1999}) \bibinfo{pages}{555
  -- 567}. \URLprefix
  \url{http://www.sciencedirect.com/science/article/pii/S0952197699000305}.
  \DOIprefix\doi{https://doi.org/10.1016/S0952-1976(99)00030-5}.
\bibitem[{Mosb{\ae}k(2014)}]{aa2621ddfa01435aaab80bdd75ff07ad}
\bibinfo{author}{R.~Mosb{\ae}k}, \bibinfo{title}{Solid Oxide Fuel Cell Stack
  Diagnostics}, Ph.D. thesis, \bibinfo{year}{2014}.
\bibitem[{Leonide(2010)}]{leonide2010sofc}
\bibinfo{author}{A.~Leonide}, \bibinfo{title}{SOFC Modelling and Parameter
  Identification by Means of Impedance Spectroscopy}, Schriften des Instituts
  f{\"u}r Werkstoffe der Elektrotechnik, Karlsruher Institut f{\"u}r
  Technologie, \bibinfo{publisher}{KIT Scientific Publ.}, \bibinfo{year}{2010}.
  \URLprefix \url{https://books.google.si/books?id=kE3Otpn2DS8C}.
\bibitem[{Wang and Blei(2019)}]{wang2019variational}
\bibinfo{author}{Y.~Wang}, \bibinfo{author}{D.~M. Blei},
  \bibinfo{title}{Variational {B}ayes under model misspecification},
  \bibinfo{year}{2019}. \href{http://arxiv.org/abs/1905.10859}{{\tt
  arXiv:1905.10859}}.
\bibitem[{{Jacob} et~al.(2018){Jacob}, {Alavi}, {Mahdi}, {Payne}, and
  {Howey}}]{7873246}
\bibinfo{author}{P.~E. {Jacob}}, \bibinfo{author}{S.~M.~M. {Alavi}},
  \bibinfo{author}{A.~{Mahdi}}, \bibinfo{author}{S.~J. {Payne}},
  \bibinfo{author}{D.~A. {Howey}},
\newblock \bibinfo{title}{{B}ayesian inference in non-markovian state-space
  models with applications to battery fractional-order systems},
\newblock \bibinfo{journal}{IEEE Transactions on Control Systems Technology}
  \bibinfo{volume}{26} (\bibinfo{year}{2018}) \bibinfo{pages}{497--506}.
\bibitem[{\v{S}m\'{i}dl and Quinn(2006)}]{Smidl2006a}
\bibinfo{author}{V.~\v{S}m\'{i}dl}, \bibinfo{author}{A.~Quinn},
  \bibinfo{title}{The Variational {B}ayes Method in Signal Processing},
  \bibinfo{publisher}{Springer-Verlag}, \bibinfo{year}{2006}.
  \DOIprefix\doi{10.1007/3-540-28820-1}.
\bibitem[{Bui et~al.(2016)Bui, Yan, and Turner}]{Bui2016}
\bibinfo{author}{T.~D. Bui}, \bibinfo{author}{J.~Yan}, \bibinfo{author}{R.~E.
  Turner},
\newblock \bibinfo{title}{A unifying framework for {G}aussian process
  pseudo-point approximations using power expectation propagation}
  (\bibinfo{year}{2016}). \href{http://arxiv.org/abs/1605.07066v3}{{\tt
  arXiv:1605.07066v3}}.
\bibitem[{Hensman et~al.(2014)Hensman, Matthews, and Ghahramani}]{Hensman2014}
\bibinfo{author}{J.~Hensman}, \bibinfo{author}{A.~Matthews},
  \bibinfo{author}{Z.~Ghahramani},
\newblock \bibinfo{title}{Scalable variational {G}aussian process
  classification}  (\bibinfo{year}{2014}).
  \href{http://arxiv.org/abs/1411.2005v1}{{\tt arXiv:1411.2005v1}}.
\bibitem[{Rezende et~al.(2014)Rezende, Mohamed, and Wierstra}]{Rezende2014}
\bibinfo{author}{D.~J. Rezende}, \bibinfo{author}{S.~Mohamed},
  \bibinfo{author}{D.~Wierstra},
\newblock \bibinfo{title}{Stochastic backpropagation and approximate inference
  in deep generative models}  (\bibinfo{year}{2014}).
  \href{http://arxiv.org/abs/1401.4082v3}{{\tt arXiv:1401.4082v3}}.
\bibitem[{Yang et~al.(2012)Yang, Xie, and Zhang}]{Yang2012}
\bibinfo{author}{Z.~Yang}, \bibinfo{author}{L.~Xie},
  \bibinfo{author}{C.~Zhang},
\newblock \bibinfo{title}{Variational {B}ayesian algorithm for quantized
  compressed sensing}  (\bibinfo{year}{2012}).
  \DOIprefix\doi{10.1109/TSP.2013.2256901}.
  \href{http://arxiv.org/abs/1203.4870v2}{{\tt arXiv:1203.4870v2}}.
\bibitem[{Oikonomou et~al.(2019)Oikonomou, Nikolopoulos, and
  Kompatsiaris}]{Oikonomou2019}
\bibinfo{author}{V.~P. Oikonomou}, \bibinfo{author}{S.~Nikolopoulos},
  \bibinfo{author}{I.~Kompatsiaris},
\newblock \bibinfo{title}{A novel compressive sensing scheme under the
  variational {B}ayesian framework},
\newblock \bibinfo{publisher}{IEEE}, \bibinfo{year}{2019}, pp.
  \bibinfo{pages}{1--5}. \DOIprefix\doi{10.23919/EUSIPCO.2019.8902704}.
\bibitem[{Gruhl and Sick(2016)}]{Gruhl2016}
\bibinfo{author}{C.~Gruhl}, \bibinfo{author}{B.~Sick},
\newblock \bibinfo{title}{Variational {B}ayesian inference for hidden markov
  models with multivariate {G}aussian output distributions}
  (\bibinfo{year}{2016}). \href{http://arxiv.org/abs/1605.08618v1}{{\tt
  arXiv:1605.08618v1}}.
\bibitem[{Panousis et~al.(2020)Panousis, Chatzis, and
  Theodoridis}]{Panousis2020}
\bibinfo{author}{K.~P. Panousis}, \bibinfo{author}{S.~Chatzis},
  \bibinfo{author}{S.~Theodoridis},
\newblock \bibinfo{title}{Variational conditional-dependence hidden markov
  models for human action recognition}  (\bibinfo{year}{2020}).
  \href{http://arxiv.org/abs/2002.05809v1}{{\tt arXiv:2002.05809v1}}.
\bibitem[{Levine(2018)}]{Levine2018}
\bibinfo{author}{S.~Levine},
\newblock \bibinfo{title}{Reinforcement learning and control as probabilistic
  inference: Tutorial and review}  (\bibinfo{year}{2018}).
  \href{http://arxiv.org/abs/1805.00909v3}{{\tt arXiv:1805.00909v3}}.
\bibitem[{Liu et~al.(2020)Liu, Qin, Zhang, Pei, Jiang, Feng, and
  Zhou}]{Liu2020}
\bibinfo{author}{Y.~Liu}, \bibinfo{author}{H.~Qin}, \bibinfo{author}{Z.~Zhang},
  \bibinfo{author}{S.~Pei}, \bibinfo{author}{Z.~Jiang},
  \bibinfo{author}{Z.~Feng}, \bibinfo{author}{J.~Zhou},
\newblock \bibinfo{title}{Probabilistic spatiotemporal wind speed forecasting
  based on a variational {B}ayesian deep learning model},
\newblock \bibinfo{journal}{Applied Energy} \bibinfo{volume}{260}
  (\bibinfo{year}{2020}) \bibinfo{pages}{114259}.
  \DOIprefix\doi{10.1016/j.apenergy.2019.114259}.
\bibitem[{Liu et~al.(2019)Liu, Qin, Zhang, Pei, Wang, Yu, Jiang, and
  Zhou}]{Liu2019}
\bibinfo{author}{Y.~Liu}, \bibinfo{author}{H.~Qin}, \bibinfo{author}{Z.~Zhang},
  \bibinfo{author}{S.~Pei}, \bibinfo{author}{C.~Wang}, \bibinfo{author}{X.~Yu},
  \bibinfo{author}{Z.~Jiang}, \bibinfo{author}{J.~Zhou},
\newblock \bibinfo{title}{Ensemble spatiotemporal forecasting of solar
  irradiation using variational {B}ayesian convolutional gate recurrent unit
  network},
\newblock \bibinfo{journal}{Applied Energy} \bibinfo{volume}{253}
  (\bibinfo{year}{2019}) \bibinfo{pages}{113596}.
  \DOIprefix\doi{10.1016/j.apenergy.2019.113596}.
\bibitem[{Choi et~al.(2018{\natexlab{a}})Choi, Kikumoto, Choudhary, and
  Ooka}]{Choi2018}
\bibinfo{author}{W.~Choi}, \bibinfo{author}{H.~Kikumoto},
  \bibinfo{author}{R.~Choudhary}, \bibinfo{author}{R.~Ooka},
\newblock \bibinfo{title}{{B}ayesian inference for thermal response test
  parameter estimation and uncertainty assessment},
\newblock \bibinfo{journal}{Applied Energy} \bibinfo{volume}{209}
  (\bibinfo{year}{2018}{\natexlab{a}}) \bibinfo{pages}{306--321}.
  \DOIprefix\doi{10.1016/j.apenergy.2017.10.034}.
\bibitem[{Choi et~al.(2018{\natexlab{b}})Choi, Menberg, Kikumoto, Heo,
  Choudhary, and Ooka}]{Choi2018a}
\bibinfo{author}{W.~Choi}, \bibinfo{author}{K.~Menberg},
  \bibinfo{author}{H.~Kikumoto}, \bibinfo{author}{Y.~Heo},
  \bibinfo{author}{R.~Choudhary}, \bibinfo{author}{R.~Ooka},
\newblock \bibinfo{title}{{B}ayesian inference of structural error in inverse
  models of thermal response tests},
\newblock \bibinfo{journal}{Applied Energy} \bibinfo{volume}{228}
  (\bibinfo{year}{2018}{\natexlab{b}}) \bibinfo{pages}{1473--1485}.
  \DOIprefix\doi{10.1016/j.apenergy.2018.06.147}.
\bibitem[{Pasquier and Marcotte(2020)}]{Pasquier2020}
\bibinfo{author}{P.~Pasquier}, \bibinfo{author}{D.~Marcotte},
\newblock \bibinfo{title}{Robust identification of volumetric heat capacity and
  analysis of thermal response tests by {B}ayesian inference with correlated
  residuals},
\newblock \bibinfo{journal}{Applied Energy} \bibinfo{volume}{261}
  (\bibinfo{year}{2020}) \bibinfo{pages}{114394}.
  \DOIprefix\doi{10.1016/j.apenergy.2019.114394}.
\bibitem[{Dolenc et~al.(2020)Dolenc, Nusev, Bo{\v{s}}koski, Morel, Mougin, and
  Juri{\v{c}}ić}]{electrimacs}
\bibinfo{author}{B.~Dolenc}, \bibinfo{author}{G.~Nusev},
  \bibinfo{author}{P.~Bo{\v{s}}koski}, \bibinfo{author}{B.~Morel},
  \bibinfo{author}{J.~Mougin}, \bibinfo{author}{{\DJ}.~Juri{\v{c}}ić},
  \bibinfo{title}{Probabilistic deconvolution of solid oxide fuel cell
  impedance spectra}, \bibinfo{year}{2020}.
  \DOIprefix\doi{10.1007/978-3-030-37161-6}.
\bibitem[{Hoffman and Gelman(2011)}]{Hoffman2011}
\bibinfo{author}{M.~D. Hoffman}, \bibinfo{author}{A.~Gelman},
\newblock \bibinfo{title}{The no-u-turn sampler: Adaptively setting path
  lengths in hamiltonian {M}onte {C}arlo}  (\bibinfo{year}{2011}).
  \href{http://arxiv.org/abs/1111.4246}{{\tt arXiv:1111.4246}}.
\bibitem[{Salvatier et~al.(2016)Salvatier, Wiecki, and
  Fonnesbeck}]{Salvatier2016}
\bibinfo{author}{J.~Salvatier}, \bibinfo{author}{T.~V. Wiecki},
  \bibinfo{author}{C.~Fonnesbeck},
\newblock \bibinfo{title}{Probabilistic programming in python using {PyMC}3},
\newblock \bibinfo{journal}{{PeerJ} Computer Science} \bibinfo{volume}{2}
  (\bibinfo{year}{2016}) \bibinfo{pages}{e55}. \URLprefix
  \url{https://doi.org/10.7717/peerj-cs.55}.
  \DOIprefix\doi{10.7717/peerj-cs.55}.
\bibitem[{Betancourt(2017)}]{Betancourt2017}
\bibinfo{author}{M.~Betancourt},
\newblock \bibinfo{title}{A conceptual introduction to {H}amiltonian {M}onte
  {C}arlo}  (\bibinfo{year}{2017}). \href{http://arxiv.org/abs/1701.02434}{{\tt
  arXiv:1701.02434}}.
\bibitem[{Blei et~al.(2016)Blei, Kucukelbir, and McAuliffe}]{Blei2016}
\bibinfo{author}{D.~M. Blei}, \bibinfo{author}{A.~Kucukelbir},
  \bibinfo{author}{J.~D. McAuliffe},
\newblock \bibinfo{title}{Variational inference: A review for statisticians}
  (\bibinfo{year}{2016}). \DOIprefix\doi{10.1080/01621459.2017.1285773}.
  \href{http://arxiv.org/abs/1601.00670v9}{{\tt arXiv:1601.00670v9}}.
\bibitem[{Paszke et~al.(2019)Paszke, Gross, Massa, Lerer, Bradbury, Chanan,
  Killeen, Lin, Gimelshein, Antiga, Desmaison, Kopf, Yang, DeVito, Raison,
  Tejani, Chilamkurthy, Steiner, Fang, Bai, and Chintala}]{NEURIPS2019_9015}
\bibinfo{author}{A.~Paszke}, \bibinfo{author}{S.~Gross},
  \bibinfo{author}{F.~Massa}, \bibinfo{author}{A.~Lerer},
  \bibinfo{author}{J.~Bradbury}, \bibinfo{author}{G.~Chanan},
  \bibinfo{author}{T.~Killeen}, \bibinfo{author}{Z.~Lin},
  \bibinfo{author}{N.~Gimelshein}, \bibinfo{author}{L.~Antiga},
  \bibinfo{author}{A.~Desmaison}, \bibinfo{author}{A.~Kopf},
  \bibinfo{author}{E.~Yang}, \bibinfo{author}{Z.~DeVito},
  \bibinfo{author}{M.~Raison}, \bibinfo{author}{A.~Tejani},
  \bibinfo{author}{S.~Chilamkurthy}, \bibinfo{author}{B.~Steiner},
  \bibinfo{author}{L.~Fang}, \bibinfo{author}{J.~Bai},
  \bibinfo{author}{S.~Chintala},
\newblock \bibinfo{title}{Pytorch: An imperative style, high-performance deep
  learning library},
\newblock in: \bibinfo{editor}{H.~Wallach}, \bibinfo{editor}{H.~Larochelle},
  \bibinfo{editor}{A.~Beygelzimer}, \bibinfo{editor}{F.~d\textquotesingle
  Alch\'{e}-Buc}, \bibinfo{editor}{E.~Fox}, \bibinfo{editor}{R.~Garnett}
  (Eds.), \bibinfo{booktitle}{Advances in Neural Information Processing Systems
  32}, \bibinfo{publisher}{Curran Associates, Inc.}, \bibinfo{year}{2019}, pp.
  \bibinfo{pages}{8024--8035}. \URLprefix
  \url{http://papers.neurips.cc/paper/9015-pytorch-an-imperative-style-high-performance-deep-learning-library.pdf}.
\bibitem[{Kingma and Ba(2014)}]{kingma2014adam}
\bibinfo{author}{D.~P. Kingma}, \bibinfo{author}{J.~Ba}, \bibinfo{title}{Adam:
  A method for stochastic optimization}, \bibinfo{year}{2014}.
  \href{http://arxiv.org/abs/1412.6980}{{\tt arXiv:1412.6980}}.
\bibitem[{Reddi et~al.(2019)Reddi, Kale, and Kumar}]{Reddi2019}
\bibinfo{author}{S.~J. Reddi}, \bibinfo{author}{S.~Kale},
  \bibinfo{author}{S.~Kumar},
\newblock \bibinfo{title}{On the convergence of adam and beyond}
  (\bibinfo{year}{2019}). \href{http://arxiv.org/abs/1904.09237v1}{{\tt
  arXiv:1904.09237v1}}.
\bibitem[{Bo{\v{s}}koski et~al.(2017)Bo{\v{s}}koski, Debenjak, and
  Boshkoska}]{Boskoski2017}
\bibinfo{author}{P.~Bo{\v{s}}koski}, \bibinfo{author}{A.~Debenjak},
  \bibinfo{author}{B.~M. Boshkoska}, \bibinfo{title}{Fast Electrochemical
  Impedance Spectroscopy}, \bibinfo{publisher}{Springer International
  Publishing}, \bibinfo{year}{2017}. \DOIprefix\doi{10.1007/978-3-319-53390-2}.
\bibitem[{Wang and Blei(2018)}]{Wang_2018}
\bibinfo{author}{Y.~Wang}, \bibinfo{author}{D.~M. Blei},
\newblock \bibinfo{title}{Frequentist consistency of variational {B}ayes},
\newblock \bibinfo{journal}{Journal of the American Statistical Association}
  \bibinfo{volume}{114} (\bibinfo{year}{2018}) \bibinfo{pages}{1147–1161}.
  \URLprefix \url{http://dx.doi.org/10.1080/01621459.2018.1473776}.
  \DOIprefix\doi{10.1080/01621459.2018.1473776}.
\bibitem[{Lang et~al.(2008)Lang, Auer, Eismann, Szabo, and
  Wagner}]{LANG20087509}
\bibinfo{author}{M.~Lang}, \bibinfo{author}{C.~Auer},
  \bibinfo{author}{A.~Eismann}, \bibinfo{author}{P.~Szabo},
  \bibinfo{author}{N.~Wagner},
\newblock \bibinfo{title}{Investigation of solid oxide fuel cell short stacks
  for mobile applications by electrochemical impedance spectroscopy},
\newblock \bibinfo{journal}{Electrochimica Acta} \bibinfo{volume}{53}
  (\bibinfo{year}{2008}) \bibinfo{pages}{7509--7513}. \URLprefix
  \url{https://www.sciencedirect.com/science/article/pii/S0013468608004982}.
  \DOIprefix\doi{https://doi.org/10.1016/j.electacta.2008.04.047},
  \bibinfo{note}{7th International Symposium on Electrochemical Impedance
  Spectroscopy}.
\bibitem[{Tenreiro(2013)}]{tenreiro2013boundary}
\bibinfo{author}{C.~Tenreiro},
\newblock \bibinfo{title}{Boundary kernels for distribution function
  estimation},
\newblock \bibinfo{journal}{REVSTAT Statistical Journal} \bibinfo{volume}{11}
  (\bibinfo{year}{2013}) \bibinfo{pages}{169--190}.

\end{thebibliography}
	
	\appendix
	\section{Supplementary material: numerical implementation}​​​​
	\label{supp_mat}​​​​
	Supplementary material containing numerical implementation of the \gls{vb} algorithm can be found at
	\url{https://repo.ijs.si/lznidaric/variational-bayes-supplementary-material}​​.
	\texttt{Python Jupyter} notebook file \texttt{SupplementaryMaterialSVI.ipynb} contains the main algorithm.
	Data set used to recreate the results presented in this article can be found in \texttt{NumPy} file \texttt{dataset.npz}.
	Repository contains all the needed scripts for implementation of \gls{vb} algorithm.
	Simulation of data was however done separately, so the code is not included.

	\section{Additional results on simulated measurements}
	\label{Additional_numerical}
	\mnewn{Reviewer 2, Comment 2.5, Comment 2.17}{
		Additional  simulation runs with the model from section 3 are performed under different noise levels in current and voltage.
		Noise $n_1(t)$ and $n_2(t)$ was added both  to the voltage $u(t)$ as well as current $i(t)$   respectively.
		It should be noted that $n_1(t)$ and $n_2(t)$ are zero-mean and uncorrelated.
		Most interesting results were selected and are presented here.
		Table~\ref{tb:add_meas} contains the information about selected measurements. The results suggest that even in the case of high noise level the \gls{vb} produces plausible estimate of the true \gls{eis} characteristic.}
	
	\begin{table}[!htbp]
		\centering
		\caption{Noise levels used in simulation.}
		\begin{tabular}{cccc} 
			\toprule[1pt]
			Measurement number &
			$1$&
			$2$&
			$3$
			\\
			\midrule
			Input noise &
			$\mathcal{N} (0.0,0.0)$&
			$\mathcal{N} (0.0,0.00005)$&
			$\mathcal{N} (0.0,0.05)$\\
			\midrule
			Output noise&
			$\mathcal{N} (0,0.0)$&
			$\mathcal{N} (0,0.001)$&
			$\mathcal{N} (0,0.05)$
			\\				
			\bottomrule[1.5pt]
		\end{tabular}
		\label{tb:add_meas}
	\end{table}	
	\begin{figure}
		\centering
		\includegraphics{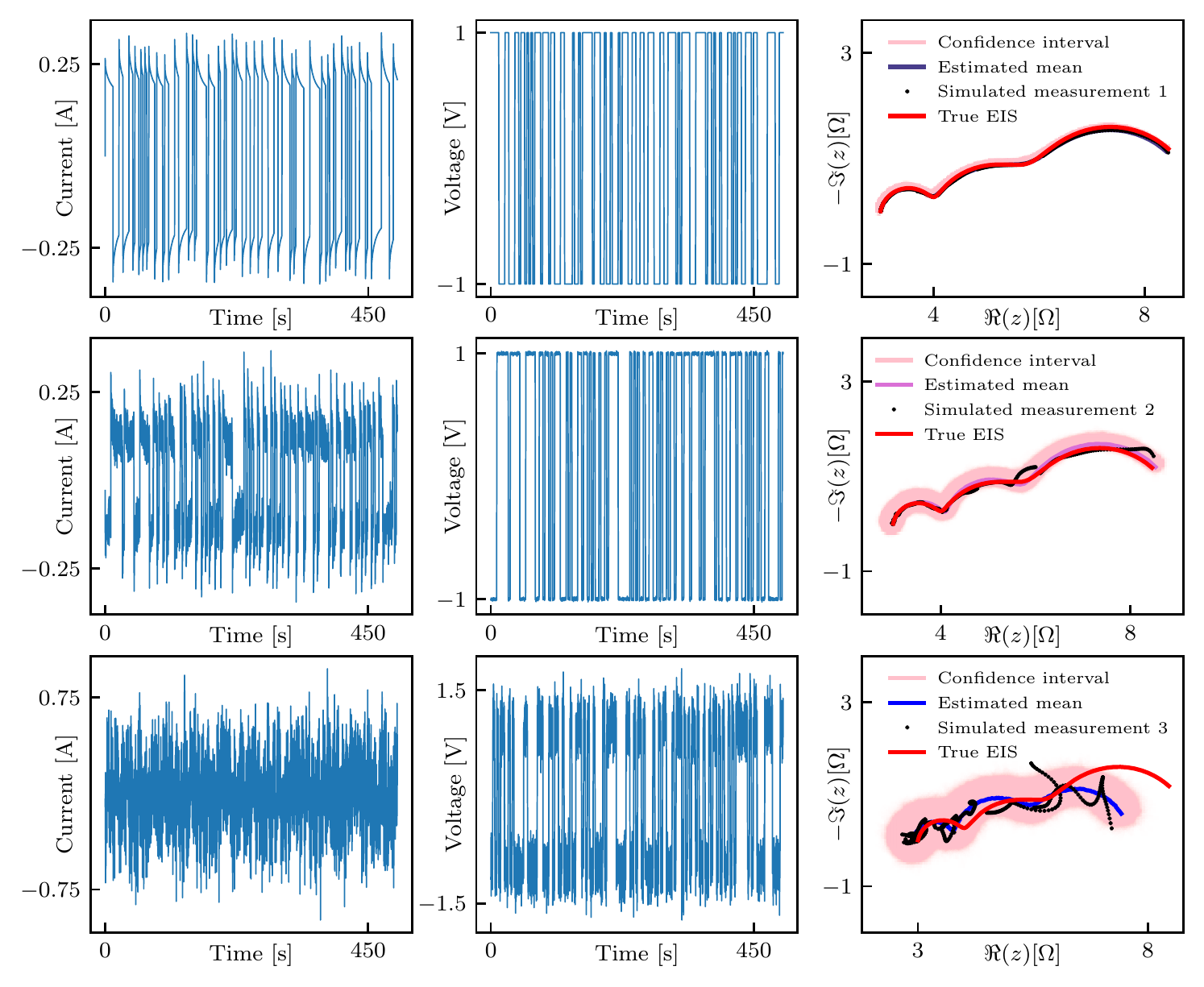}
		\caption{Simulated current, voltage and the resulting \gls{eis} curves with different noise settings along with \gls{vb} estimated mean and the confidence region.}
		\label{fig:toyexamplethreeeis}
	\end{figure}
	
	\begin{figure}
		\centering
		\includegraphics{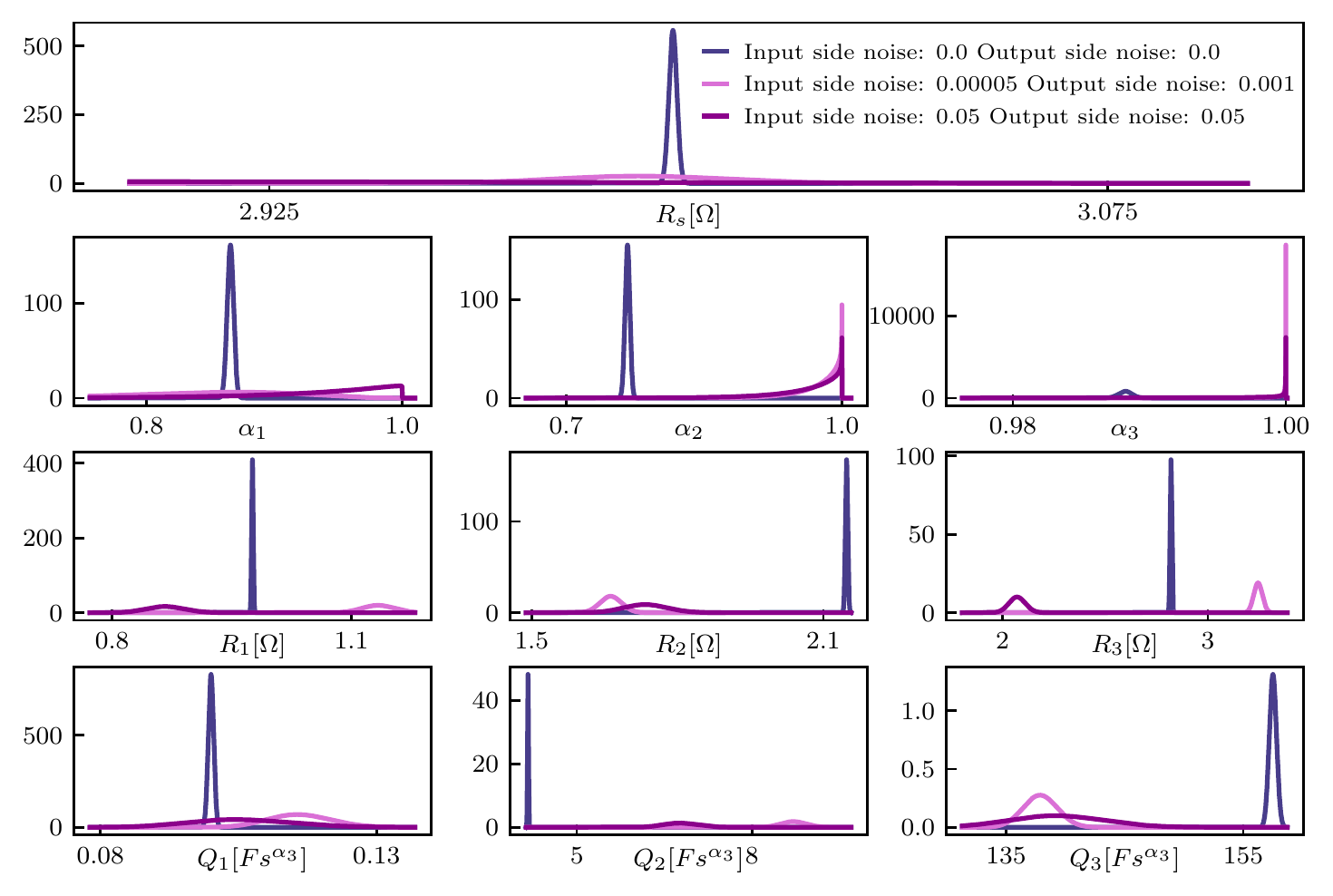}
		\caption{Posterior parameter distributions obtained by  \gls{vb} approach on  simulated noisy data. Uncertainty in the parameters increases with increasing level of noise used for simulation.}
		\label{fig:toyexamplethreepost}
	\end{figure}
	
	\section{Additional results on experimental measurements}
	\label{Additional_experimental}
\begin{figure}
		\centering
		\includegraphics{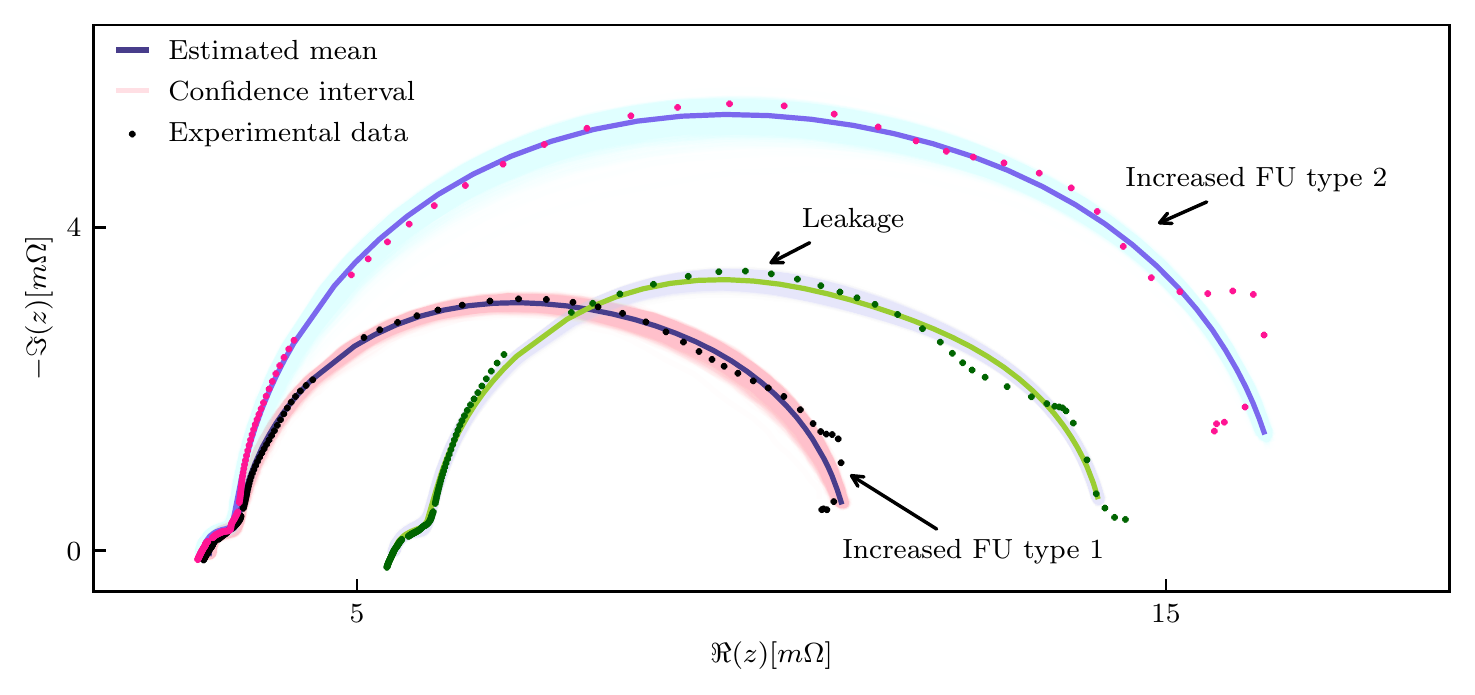}
		\caption{\gls{vb} results for 3 \gls{eis} measurements at different operational conditions.}
		\label{fig:differentfueis}
	\end{figure}

\begin{figure}
		\centering
		\includegraphics{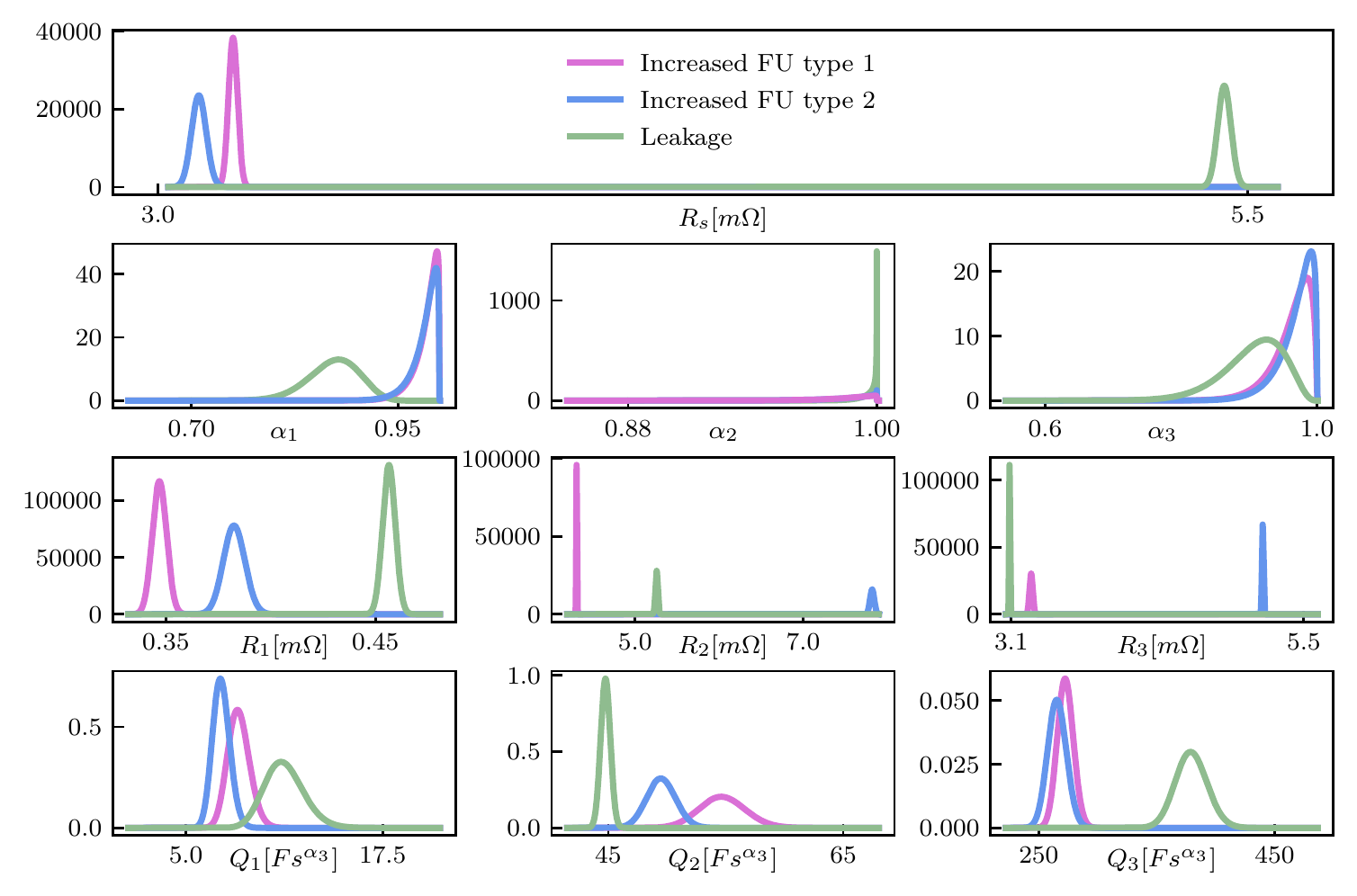}
		\caption{Posterior distributions of ECM parameters obtained on 3  measurements under different conditions explained in Table \ref{tb:exper_added}.}
		\label{fig:differentfupost}
	\end{figure}

	\mnew{Reviewer 2, Comment 2.4}{The method was tested on a  set of measured \gls{eis} data at different currents in healthy state and on data under leakage fault.
	Measurement presented in the main article is obtained during the nominal running conditions of the \gls{sofc} which are 77\% fuel utilization and current of 32 A.
	Main operating conditions of additional measurements are presented in \tablename~\ref{tb:exper_added}.
	Measurement 1 and Measurement 2 are obtained during different fuel utilization settings.
	For Measurement 1, fuel utilization was increased from 77\% to 72\% by decreasing the gas flow and keeping the current steady at ~32 A.
	On the flip side, fuel utilization in Measurement 2 was increased from 72\% to 87\% by increasing the electrical current from ~32 A to 36.4 A. 
	Measurement 3 was recorded after a leakage fault occured in the plant.
	In \figurename~\ref{fig:differentfueis} the results of \gls{vb} algorithm are presented.
	The \gls{vb} algorithm returns good estimates for each of the measurement, irrespective of the fact that the same set of variational distributions are applied.
	The resulting posterior distributions for each parameter are presented in \figurename~\ref{fig:differentfupost}.
	}

	\begin{table}[!htbp]
		\centering
		\caption{Operational conditions for the measurements.}
		\begin{tabular}{ccccccc} 
			\toprule[1pt]
			Measurement number &
			$1$&
			$2$&
			$3$
			\\
			\midrule
			Current [A] &
			
			$32.21$&
			$36.39$&
			$32.04$	
			\\
			\midrule
			Fuel utilization [$\%$]&
			$82$&
			$87$&
			$77$
			\\				
			\bottomrule[1.5pt]
		\end{tabular}
		\label{tb:exper_added}
	\end{table}
	
\end{document}